\documentclass[doublespacing]{elsart}
\usepackage{amsmath,mathrsfs,graphicx,epsfig,amssymb}
\def\pmb#1{\setbox0=\hbox{$#1$}\kern-0.015em\copy0\kern-\wd0%
\kern0.03em\copy0\kern-\wd0\kern-0.015em\raise0.03em\box0}
\begin{document}
\runauthor{Yee Yee Oo et al.}
\begin{frontmatter}
\title{Scattering Polarization in the Presence of Magnetic and 
Electric 
Fields}
\author[Mdy]{Yee Yee Oo}
\author[Pisa]{M. Sampoorna\thanksref{someone}}
\author[Pisa]{K. N. Nagendra}
\author[Rome]{Sharath Ananthamurthy}
\author[Pisa]{G. Ramachandran\corauthref{cor}}
\ead{gr@iiap.res.in}
\address[Mdy]{Department of Physics, Mandalay University, Mandalay,
Myanmar}
\address[Pisa]{Indian Institute of Astrophysics, Bangalore 560-034, 
India}
\address[Rome]{Department of Physics, Bangalore University, 
Bangalore 560-056,
India}
\corauth[cor]{Indian Institute of Astrophysics, Bangalore 560-034, 
India}
\thanks[someone]{Joint Astronomy Program, Dept. of Physics, IISc,
  Bangalore-560 012, India.}
\begin{abstract}
The polarization of radiation by scattering on an atom embedded in 
combined external quadrupole electric and uniform magnetic fields is 
studied theoretically. Limiting cases of scattering under Zeeman 
effect and 
Hanle effect in weak magnetic fields are discussed. The theory is 
general 
enough to handle scattering in intermediate magnetic fields 
(Hanle-Zeeman 
effect) and for arbitrary orientation of magnetic field. The 
quadrupolar 
electric field produces asymmetric line shifts and causes 
interesting 
level-crossing phenomena either in the absence of an ambient 
magnetic field
or in its presence. It is shown that the quadrupolar electric field 
produces 
an additional depolarization in the $Q/I$ profiles and rotation of 
the plane of 
polarization in the $U/I$ profile over and above that arising from 
magnetic 
field itself. This characteristic may have a diagnostic potential to 
detect 
steady state and time varying electric fields that surround 
radiating atoms in 
Solar atmospheric layers. 
\end{abstract}
\end{frontmatter}
\section{Introduction}
Scattering of polarized radiation by an atom is a topic of 
considerable
interest to astrophysics ever since Hale~\cite{hale} first observed 
polarization related to Zeeman effect in spectral lines originating 
in Sun spots. The polarized radiation is usually expressed in terms 
of the Stokes parameters. The concept of scattering
matrix connecting the Stokes vector ${\bf  {S}}'$
of incident radiation to the Stokes vector ${\bf{S}}$
of scattered radiation was introduced quite early in the context of 
Rayleigh scattering~\cite{chandra}. Polarized radiation in spectral 
lines formed in the presence 
of an external magnetic field has been studied widely and a 
comprehensive theoretical framework
has been 
developed~\cite{landi-1,landi-2,landi-3,landi-4,lbs1,lbs2,lbs3,bom-1
,bom-2,stenflo2,stenflo3,stenflo4,bs99}. 
The Hanle effect is a depolarizing phenomenon which arises due to 
`partially overlapping' magnetic substates in the presence of weak 
magnetic fields, when the splitting produced is of the same order or 
less than the natural widths. Favati et al.~\cite{favati} proposed 
the name `second Hanle effect' for a similar effect in 
`electrostatic fields'. Casini and Landi 
Degl'Innocenti~\cite{casini} have discussed
the problem in the presence of electric and magnetic fields for the 
particular 
case of hydrogen Lyman $\alpha$ line. It was followed by a more 
recent paper by 
Casini~\cite{casini1}. The relative contributions of static 
external electric fields, motional electric fields and magnetic 
fields in 
the case of hydrogen Balmer lines, have been studied
by Brillant et al.~\cite{brillant}. A historical perspective and 
extensive 
references to earlier literature on polarized line scattering can be 
found in Stenflo~\cite{stenflo2}, Trujillo Bueno et al.~\cite{tru} 
and 
Landi Degl'Innocenti and Landolfi~\cite{ll}. 

A quantum electrodynamic theory of Hanle-Zeeman redistribution 
matrices has
been developed by Bommier \cite{bom-1,bom-2} and Landi 
Degl'Innocenti \& 
Co-workers (see the book by \cite{ll}). The formulation presented in 
\cite{bom-1} and \cite{bom-2} includes the effects of partial 
frequency 
redistribution (PRD) in line scattering for a two-level atom. It is 
a 
perturbation theory, in which PRD effects appear in the fourth 
order. The 
theory presented in \cite{ll} and references cited therein, 
considers only
complete frequency redistribution (CRD) in line scattering.

A classical theory of line scattering PRD for the Hanle-Zeeman 
effect has been
formulated by Bommier \& Stenflo \cite{bs99}. This theory is 
non-perturbative 
and describes the scattering process in a transparent way. The 
classical 
theory for Hanle-Zeeman scattering developed by Stenflo 
\cite{stenflo4} 
considered only coherent scattering in the laboratory frame. In 
\cite{bs99} 
the redistribution matrices were derived in the atomic rest frame. 
The
corresponding laboratory frame redistribution matrices have been 
derived
in \cite{sam1a}. The equivalence between the classical 
(non-perturbative 
theory) and quantum electrodynamic (perturbative theory) 
redistribution 
matrices for the triplet case is established in \cite{sam1}. In all 
these 
papers only the dipole type line scattering transitions in the 
presence of 
pure magnetic fields is considered. Taking into account all higher 
order 
multipoles as well, polarization of line radiation in the 
presence of external electric quadrupole and uniform magnetic fields 
was studied 
\cite{Oo1,Oo2}, where scattering of radiation by atoms, however, was 
not considered. 

The purpose of the present paper is to develop a quantum 
electrodynamical approach to scattering processes in the presence of 
external electric and magnetic fields of `arbitrary strengths', 
taking also into consideration all other multipole type transitions 
apart from the usually dominant electric dipole transition. The 
atomic electron is represented using non-relativistic quantum theory 
including spin. The radiation field is described in terms of its 
electric and magnetic multipole
states, in a second quantized formalism. The external electric field 
is assumed to be `quadrupolar' in nature, while the magnetic field 
is uniform and arbitrarily oriented with reference to the principal 
axes frame of the electric quadrupole field. This general formalism 
can be employed also to 
solve the scattering problems involving linear steady state electric 
fields 
at the radiating atom. 

In Sect.~2 we describe the theoretical formulation. In Sect.~3 the 
scattering 
matrix for the general physical situation is derived. The particular 
case of the dipole transitions for a triplet is also considered, for 
the 
purpose of comparison with Stenflo \cite{stenflo4} in the pure 
magnetic field 
limit. Sect.~4 contains numerical results and discussions. 
Conclusions are 
presented in Sect.~5.

\section{Theoretical formalism for scattering}
\begin{figure}
\begin{center}
\includegraphics[width=8cm,height=8cm]{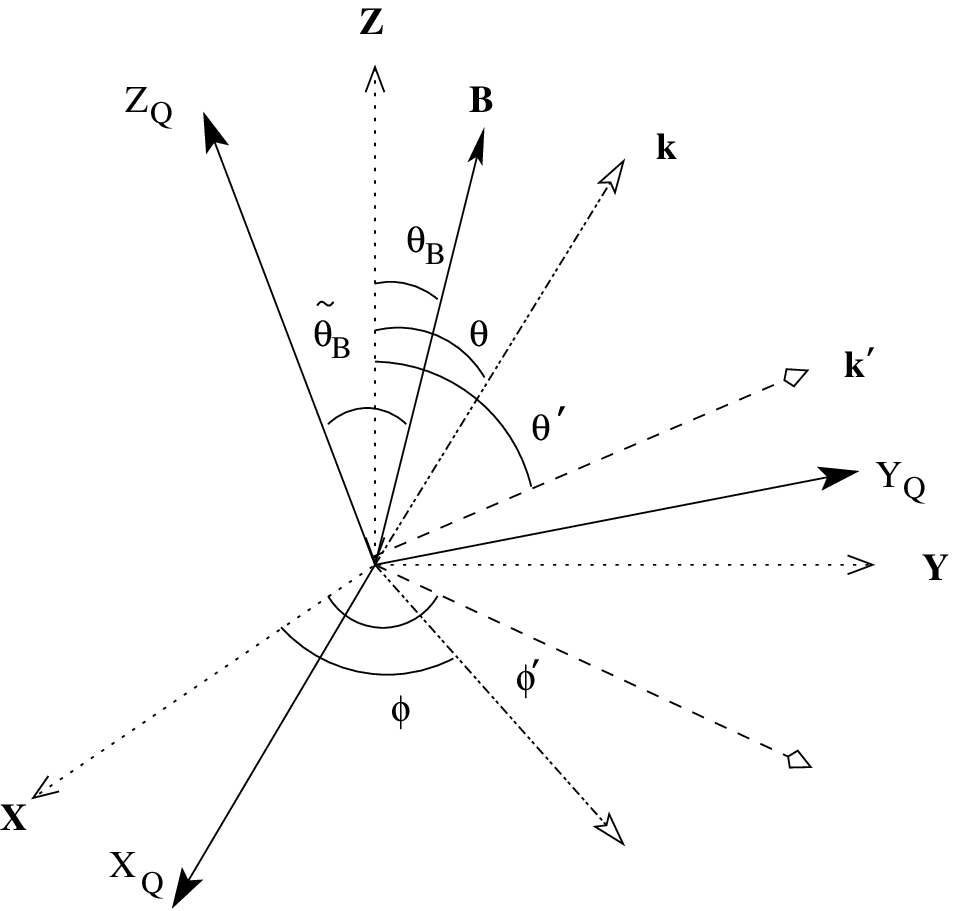}
\end{center}
\caption{ The scattering geometry:
$({\rm X}_{Q}, {\rm Y}_{Q}, {\rm Z}_{Q})$ refers to the Principal
Axes Frame (PAF) characterizing the electric quadrupole field.
The radiation is incident along $(\theta', \phi')$ and scattered 
along
$(\theta, \phi)$ with respect to the astrophysical reference frame 
(ARF) 
denoted by $({\rm X}, {\rm Y}, {\rm Z})$. The magnetic field
${\vec B}$ is oriented along $(\widetilde \theta_{B}, \widetilde 
\phi_{B})$
with reference to PAF and $(\theta_{B}, \phi_{B})$ with reference to 
the
astrophysical reference frame (the azimuthal angles $\widetilde 
\phi_{B}$
and $\phi_{B}$ are not marked in the figure). }
\label{geometry}
\end{figure}

We consider polarized radiation incident on an atom along an 
arbitrary 
direction $(\theta',\phi')$ and getting
scattered into a direction $(\theta, \phi)$ with respect to a 
conveniently 
chosen right handed Cartesian coordinate
system, referred to as the Astrophysical Reference Frame (ARF) and 
shown as 
$(X,Y,Z)$ in Fig.~\ref{geometry}. If $\nu'$ and $\nu$ denote
respectively the frequencies of the incident and scattered 
radiation, we 
may define wave vectors ${\bf k}'$ and ${\bf k}$ with 
polar co-ordinates $(k',\theta',\phi')$ and $(k,\theta,\phi)$ where 
$k'=2\pi\nu'=\omega'$ and $k=2\pi\nu=\omega$ in
natural units with $\hbar=1, c=1$ and mass of the electron $m_e=1$. 
The atom is 
exposed to an external magnetic field 
${\bf B}$ with strength $B$, directed along $(\theta_B, \phi_B)$ and 
an 
electric quadrupole field characterized by 
strength $A$ and asymmetry parameter $\eta$ in its Principal Axes 
Frame 
(PAF), which is denoted by $(X_Q, Y_Q, Z_Q)$ 
in Fig.~\ref{geometry}. The transformation to PAF from ARF is 
achieved by a 
rotation $R(\alpha_Q, \beta_Q, \gamma_Q)$ through Euler angles 
$(\alpha_Q, \beta_Q, \gamma_Q)$ as defined by Rose~\cite{rose}. 
The magnetic field 
${\bf B}$ is directed along 
$(\widetilde\theta_{B},\widetilde\phi_{B})$ with respect to PAF. 
Following Rose~\cite{rose}, we define left and right circular states 
of 
polarization ${\hat {\pmb \varepsilon}}_{\mu}$ ($\mu=\pm 1$) 
respectively, which 
are mutually orthogonal to each other and to 
${\bf k}$. We use here the symbol ${\hat {\pmb 
\varepsilon}}_{\mu=\pm 1}$ instead 
of ${\hat {\bf u}}_{{p}=\pm 1}$ employed in Rose~\cite{rose}. Like 
wise 
${\hat {\pmb \varepsilon}}'_{\mu'=\pm 1}$, which are orthogonal to 
${\bf k}'$.
Any arbitrary state of polarization ${\hat 
{\pmb \varepsilon}}^{'}$ of the incident radiation may then be 
expressed as ${\hat 
{\pmb \varepsilon}}^{'}=c'_{+1}\ {\hat 
{\pmb \varepsilon}}^{'}_{+1}+c'_{-1}\ {\hat {\pmb 
\varepsilon}}^{'}_{-1}$ using 
appropriate coefficients $c'_{\pm 1}$, which are in general 
complex and satisfy $|c'_{+1}|^{2}+|c'_{-1}|^{2}=1$. We, therefore, 
denote 
the orthonormal states of polarized
incident radiation by $|{\bf k}', \mu'\rangle$, with $\mu'=\pm 1$. 
We seek the 
probability for scattering into two
polarized states of scattered radiation $|{\bf k}, \mu\rangle$, 
$\mu=\pm 1$ 
on an atom which is initially in a
state $\psi_{i}$ with energy $E_{i}$ before scattering and makes a 
transition to a final state $\psi_{f}$ with energy
$E_{f}$, in the process of scattering of polarized radiation. 

\subsection{Energy levels of an atom in electric quadrupole and 
uniform 
magnetic fields}
The energy levels of an electron in an atom are primarily determined 
by the 
Hamiltonian 
\begin{equation}
H_{0}= -\frac{1}{2}\bigtriangledown^{2}+V(r)\ , \label{unperturbed}
\end{equation}
where $V(r)$ denotes its Coulomb interaction  with the nucleus. If 
we
start with the Dirac equation~\cite{casini} and use its 
non-relativistic 
reduction, terms like
spin-orbit interaction may also be included in $H_{0}$. In the 
absence of 
external fields, the energy levels of the atom are determined by 
\begin{equation}
H^{A}_{0}= \sum^{Z}_{i=1} H_{0}(i)+\sum^{Z}_{i>j=1} 
\frac{e^2}{r_{ij}}\ ,
\end{equation}
where $e$ denotes the charge of the electron, $r_{ij}=|{\bf 
r}_{i}-{\bf 
r}_{j}|$ and $Z$ denotes the atomic number. If $E$ denotes an energy 
level
and $\psi$ the corresponding wave function of the atom with total 
angular
momentum $J$, it is well known in the context of Zeeman effect that 
$E$
gets split into $(2J+1)$ equally spaced levels $E_{M}=E+gBM$ with 
corresponding energy eigenstates $|JM\rangle_{\bf B}$, 
$M=J, J-1, \cdots,-J+1,-J$, when the atom is exposed to an external 
uniform
magnetic field
${\bf B}$ with strength $B$. The states $|JM\rangle_{\bf B}$ are 
defined
with the axis of quantization chosen along ${\bf B}$ and $g$ denotes 
the 
magnetic-gyro ratio or 
Land\'e $g$-factor. For $B<100$ gauss, when $gB$ is of the same 
order
as the width of a line, Hanle effect~\cite{hanle} takes place. For a 
line in 
the optical range, the region
$100<B<1000$ gauss is generally referred to as the Hanle-Zeeman 
regime.
For $B<1$ gauss, one has to pay attention to the interaction of 
electron with the magnetic and electric moments of the nucleus, 
which 
give rise to hyperfine splitting~\cite{BJ,shore}. 
If the atom is exposed to an external
electric quadrupole field either by itself or in combination with 
${\bf B}$, the splitting of the energy levels is not, in general, 
equally
spaced~\cite{muha1,muha2,Oo1,Oo2} and in such scenarios, the 
atomic Hamiltonian in PAF is given by 
\begin{equation}
H_{A}=H_{0}^{A}+g \ {\bf J} \cdot {\bf B}+ A \Big[ 
2J^{2}_{z}-J^{2}_{x}-
J^{2}_{y} + \eta (J^{2}_{x}-J^{2}_{y}) \Big]\ .
\end{equation}
The split energy levels may be denoted by $E_{s}$, where
$s$ takes values $s=1, 2, \cdots, (2J+1)$ starting
from the lowest level ($s=1$) for a given $J$. The corresponding
energy eigenstates may be denoted by $|J,s\rangle$, which are 
expressible
as
\begin{equation}
|J,s\rangle=\sum^{J}_{M=-J} a^{s}_{M} (A,\eta,B,\widetilde 
\theta_{B},
\widetilde \phi_{B})\ |JM\rangle_{Q}, \quad s=1,2,\cdots,(2J+1) 
\label{general}
\end{equation}
where $|JM\rangle_{Q}$ are defined with the quantization axis chosen
along the $Z$-axis, $Z_{Q}$ of the $PAF$. The notation $c^{i}_{m_u}$
was used in~\cite{Oo2} for $J=1,3/2$ to denote the expansion 
coefficients,
without any specified convention for ordering of the levels. We
may rewrite Eq.~(\ref{general}) as
\begin{equation}
|J,s\rangle=\sum^{J}_{m=-J}\ c^{s}_{m}\ |Jm\rangle\ , 
\label{without}
\end{equation}
in terms of the $|Jm\rangle$ states, which are defined with respect 
to
the $Z$-axis of ARF chosen as the quantization axis. Clearly,
\begin{equation}
c^{s}_{m} = \sum^{J}_{M=-J} D^{J}_{mM}(\alpha_{Q}, \beta_{Q}, 
\gamma_{Q})
\ a^{s}_{M}(A, \eta, B, \widetilde \theta_{B}, \widetilde \phi_{B})\ 
,
\label{coefficient}
\end{equation}
and hence $c^{s}_{m}$ depend on $\alpha_{Q}, \beta_{Q}, \gamma_{Q}, 
A,
\eta, {\bf B}$. If the magnetic field is absent, the $c^{s}_{m}$ 
depend
only on $\alpha_{Q}, \beta_{Q}, \gamma_{Q}, A, \eta$ since 
$a^{s}_{M}$ in
that case~\cite{Oo1} depend only on $A$ and $\eta$. 
It may be noted that the 
frame of reference employed in~\cite{Oo1,Oo2} was $PAF$ itself, 
i.e., 
$\alpha_{Q}, \beta_{Q}, \gamma_{Q}=0$; hence
$\theta_{B}, \phi_{B}$ was used there instead of the 
$\widetilde \theta_{B}, \widetilde \phi_{B}$ here and the Euler 
angles
$(\alpha_{Q}, \beta_{Q}, \gamma_{Q})$ find no mention there. On the 
other
hand, if the electric quadrupole field is absent and the atom is 
exposed
only to a magnetic field ${\bf B}$ directed along $(\theta_{B}, 
\phi_{B})$,
it is clear that 
\begin{equation}
c^{s}_{m}= \sum_{M=-J}^{J}D^{1}_{mM}(\phi_{B}, \theta_{B}, 0)\ 
\delta_{M,s-J-1}\ ,
\label{special}
\end{equation}
which reduces to
\begin{equation}
c^{s}_{m}= \delta_{s,J+m+1}\ , \label{simple}
\end{equation}
if the field ${\bf B}$ is along the $Z$-axis of ARF itself. 

In general, therefore, when the energy levels of an atom are defined
through $H_{A}\psi_{n}=E_{n}\psi_{n}$, the atomic wave functions 
$\psi_{n}$
are of the form given by Eq.~(\ref{without}), which specialize 
appropriately to 
$|JM\rangle_{\bf B}$ or $|Jm\rangle$ if Eq.~(\ref{special}) or 
Eq.~(\ref{simple}) are used instead of Eq.~(\ref{coefficient}). 
Thus, in general, the complete set of orthonormal energy
eigenstates of an atom in a combined external electric quadrupole
and uniform magnetic field environment may be denoted by 
$\{\psi_n\}$,
where $n$ is used as a collective index, which includes the serial 
number
$s_n$ along with the total angular momentum $J_{n}$ and all other 
quantum
numbers which may be needed to specify each $\psi_n$ uniquely. In 
the
presence of a pure magnetic field ${\bf B}$, the magnetic quantum 
number
$M_n$ replaces $s_n$ through the $\delta$-function in 
Eq.~(\ref{special}).
Moreover,
if ${\bf B}$ is along the $Z$-axis of ARF itself, $s_n$ gets 
replaced
by $m_n$ through the $\delta$-function in Eq.~(\ref{simple}).

In general, a summation over $n$ as in $\sum_{n}|\psi_n\rangle 
\langle \psi_n|=1$, implies a summation with respect to $s_n$ as 
well.
This summation over $s_n$ may be replaced by a summation with 
respect to
$M_n$ or $m_n$ in some particular cases as mentioned
above. The initial and final states of the atom before and after 
scattering
are denoted by $\psi_{i}$ and $\psi_{f}$. They also belong to $\{ 
\psi_n\}$.
We use the short hand notation
\begin{equation}
|i\rangle = |\psi_{i};{\bf k}', \mu' \rangle ; \hspace{1cm}
|f\rangle = |\psi_{f};{\bf k}, \mu \rangle \ . \label{shorthand}
\end{equation}

\subsection{Interaction of atom with the radiation field}
It is well known that the local minimal coupling i.e, 
${\bar \psi}\gamma_{\nu}\psi A_{\nu}$ (with implied summation over 
$\nu$)
of the Dirac field $\psi$ and the electromagnetic field represented 
by
the four potential $A_{\nu}, \nu=1,\cdots,4$ is the fundamental 
interaction responsible for all electrodynamical process involving 
photons 
and electrons~\cite{jauch,hamilton}. In the interaction 
representation, $\psi$ and 
$A_{\nu}$ satisfy the free field equations of Dirac and Maxwell 
respectively. The quantity 
${\bar \psi}=\psi^{\dag}\gamma_{4}$, where $\psi^{\dag}$ denotes 
the hermitian
conjugate of $\psi$ and $\gamma_1, \gamma_2, \gamma_3, \gamma_4$ are
$4 \times 4$ Dirac matrices. To facilitate calculations using the 
atomic
wave functions $\{ \psi_n \}$, we may use the non-relativistic two 
componental forms of $\psi$ and ${\bar \psi}$ 
in $c$ number theory for electrons, retain the Maxwell field in q 
number theory and represent the
interaction of the radiation field in the Coulomb gauge with the 
atom as
\begin{equation}
H_{int} = e^{i H_{A} t} \sum^{Z}_{j=1} e\ \Bigl[ - i {\bf A}
({\bf  r}_{j}, t) \cdot {\pmb {\bigtriangledown}}_{j} + 
{\frac{1}{2}}
\,{\pmb \sigma}_{j} \cdot \Bigl({\pmb \bigtriangledown}_{j}
\times {\bf A}({\bf r}_{j},t) \Bigr) \Bigr] e^{-i H_{A} t}\ , 
\label{hint}
\end{equation}
where ${\pmb \sigma}_{j}$ denote the Pauli spin matrices of the 
electron labeled $j$ located at ${\bf r}_{j}$ and $Z$ denotes the 
atomic 
number. The quantum field variable ${\bf A}({\bf r},t) $ in 
interaction 
representation may be expressed as
\begin{eqnarray}
{\bf A}({\bf r}_{j},t)= \frac{1}{(2\pi)^{3/2}} \int \frac{d^{3}k''}
{\sqrt{2\omega''}}\  \sum_{\mu''} & \Bigl[& a_{{\bf k}'' \mu''} 
{\bf A}_{{\bf k}'' \mu''}({\bf r}) e^{-i\omega'' t} + \nonumber \\
& & a^{+}_{{\bf k}'' \mu''} {\bf A}_{{\bf k}'' \mu''}({\bf r})^{*}
e^{i\omega'' t} \Bigr]\ , \label{field}
\end{eqnarray}
where $\omega''=|{\bf k}''|$ and the creation and annihilation 
operators, 
denoted by
$a^{+}_{{\bf k}\mu}$ and  $a_{{\bf k}\mu}$ respectively, satisfy the 
commutation relation
\begin{equation}
\Bigl[ a_{{\bf k} \mu} , a^{+}_{{\bf k}' \mu'}\Bigr] =
\delta({\bf k}-{\bf k}')\ \delta_{\mu \mu'}\ , \label{commutator}
\end{equation}
for any pair ${\bf k},\mu$ and ${\bf k}',\mu'$ in general, while
\begin{equation}
{\bf A}_{{\bf k} \mu}({\bf r})= {\hat {\pmb \varepsilon}}_{\mu} 
\ e^{i {\bf k} \cdot {\bf r}}\ , \label{kmu}
\end{equation}
denotes a $c$ number and ${\bf A}_{{\bf k}\mu}({\bf r})^{*}$ denotes 
its 
complex conjugate. In particular, the operators are also used to 
generate 
the initial and final states of radiation in Eq.~(\ref{shorthand}) 
through
\begin{equation}
|{\bf k}' \mu' \rangle = a^{+}_{{\bf k}' \mu'} |\ \rangle_{0} ;
\hspace{1cm} \langle {\bf k} \mu| =\  _0\langle \ | a_{{\bf k} \mu} 
\ ,
\end{equation}
where $|\ \rangle_{0}$ denotes the vacuum state of the radiation 
field.
\subsection{The Scattering process}
The $S$-matrix for scattering may be defined, as 
usual~\cite{eugen}, by 
\begin{equation}
S = {\rm lim}_{t \rightarrow \infty \atop t_{0} \rightarrow -\infty}
 U(t,t_{0})\ ,
\end{equation}
where the evolution operator satisfies
\begin{equation}
U(t,t_{0})= 1 - i \int^{t}_{t_0} dt' H_{int}(t') U(t',t_0)\ ,
\end{equation}
which on iteration leads to the perturbation series
\begin{equation}
S=1+\sum^{\infty}_{N=1} (-i)^{N} \int^{\infty}_{-\infty} dt_{1}
\int^{t_1}_{-\infty} dt_{2} \cdots \int^{t_{N-1}}_{-\infty} dt_{N}
H_{int}(t_{1}) \cdots H_{int}(t_{N}) \ ,
\end{equation}
since $H_{int}(t)$ given by Eq.~(\ref{hint}) is linear in ${\bf A}$ 
(see Eq.~(\ref{field})), the first order
$(N=1)$ term can contribute to either absorption through the first 
term in 
Eq.~(\ref{field}) or to emission through the second term in 
Eq.~(\ref{field}) and the integral over $dt_1$, from 
$-\infty \rightarrow \infty$
leads to the respective energy conservation criteria of Bohr. In the 
scattering problem under consideration,
the lowest order (in $e$) contribution to $\langle f|S|i\rangle$ is 
obtained from the $N=2$ term, which we
may denote as $\langle f|S^{2}|i\rangle$. We introduce 
$\sum_{n}|\psi_{n}\rangle\langle \psi_{n}|=1$ between
$H_{int}(t_1)$ and $H_{int}(t_2)$, neglect contribution from two 
photons in the intermediate state and employ
the notation $|n\rangle=|\psi_{n}\rangle|\ \rangle_{0}$. This leads, 
on using Eqs.~(\ref{field}) and (\ref{commutator}), to
\begin{equation}
\langle n|H_{int}(t_{2})|i \rangle =  {\mathcal A}_{ni}
({\bf k}', \mu') e^{[i(E_{n}-E_{i}-\omega')t_{2}]}\ , \label{ni}
\end{equation}
\begin{equation}
\langle f|H_{int}(t_{1})|n \rangle ={\mathcal E}_{fn}
({\bf k}, \mu) e^{[i(E_{f}+\omega -E_{n})t_{1}]}\ , \label{fn}
\end{equation}
where ${\mathcal A}_{ni}({\bf k}',\mu')$ and ${\mathcal 
E}_{f,n}({\bf k},\mu)$ denote amplitudes for
absorption and emission, involving ${\bf A}_{{\bf k}' \mu'}({\bf 
r}_{j})$ and
${\bf A}_{{\bf k} \mu}({\bf r}_{j})^{*}$ respectively, which are 
independent 
of time variable, instead of ${\bf A}({\bf r}_{j},t)$. 
We may change the variable of integration from $t_{2}$ to 
$t'_{2}=t_{2}-t_{1}$, ranging from $-\infty \rightarrow 0$, 
associate a width $\Gamma_{n}$ with $\psi_{n}$ by introducing a 
factor 
$exp(\Gamma_{n}t'_{2})$ (see \cite{weisskopf1,weisskopf2}) and 
obtain 
after completing both the integrations, the expression
\begin{equation}
\langle f|S^{2}|i \rangle = - 2\ \pi\ i\ 
\delta(E_{f}+\omega-E_{i}-\omega')
\ T_{fi}({\bf k}\mu; {\bf k}' \mu')\ , \label{smatrix}
\end{equation}
where the on-energy-shell $T$-matrix element is of the form
\begin{equation}
T_{fi}({\bf k},\mu;{\bf k}',\mu') = \sum_{n} {\mathcal E}_{fn}
({\bf k},\mu)\ \phi_{n}\ {\mathcal A}_{ni}({\bf k}',\mu') \ ,
\label{ephia}
\end{equation}
and the profile function is given by
\begin{equation}
\label{prof-fun}
\phi_{n}= (\omega_{nf}-\omega-i\Gamma_{n})^{-1}; \hspace{1cm}
\omega_{nf}=E_{n}-E_{f}\ ,
\end{equation}
on making use of 
$E_{n}-E_{i}-\omega'=\omega_{ni}-\omega'=\omega_{nf}-\omega$ by 
virtue of the energy
$\delta$-function in Eq.~(\ref{smatrix}). Using Eq.~(\ref{general}) 
and observing that $\{ \psi_{n} \}$ are completely
antisymmetric with respect to the labels $1, 2, \cdots, j, \cdots, 
Z$ of the electrons, we have
\begin{eqnarray}
{\mathcal A}_{ni}({\bf k}',\mu') &=& 
\sum_{m'_{n},m'_{i}}c^{s_{n}^{*}}_{m'_{n}}\ c^{s_{i}}_{m'_{i}}
\langle J_{n}m'_{n}|{\mathcal A}({\bf k}',\mu')|J_{i}m'_{i}\rangle 
\, ,
\nonumber \\
{\mathcal E}_{fn}({\bf k},\mu) &=& 
\sum_{m'_{f},m''_{n}}c^{s_{f}^{*}}_{m'_{f}}c^{s_{n}}_{m''_{n}}\langle
e J_{f}m'_{f}|{\mathcal E}({\bf k},\mu)|J_{n}m''_{n} \rangle \, , 
\end{eqnarray}
where the matrix elements on the right hand side satisfy
\begin{equation}
\langle J_{u}M_{u}|{\mathcal A}({\bf k},\mu)|J_{l}m_{l} 
\rangle=\langle J_{l}m_{l}|{\mathcal E}({\bf 
k},\mu)|J_{u}m_{u}\rangle^{*}\ ,
\label{absorptiona}
\end{equation}
between any pair of lower and upper atomic states and 
\begin{equation}
{\mathcal E}({\bf k},\mu)= \frac{Ze}{(2\pi)^{3/2}{\sqrt {2\omega}}}
\Bigl[ -i{\bf A}^{*}_{{\bf k},\mu} \cdot {\pmb \bigtriangledown}+
\frac{1}{2} {\pmb \sigma} \cdot ({\pmb \bigtriangledown} \times 
{\bf A}^{*}_{{\bf k},\mu}) \Bigr]\ , \label{emissiona}
\end{equation}
with respect to an electron in the atom. Since atomic transitions 
during 
absorption and emission conserve total
angular momentum and parity, we use the standard multipole 
expansion~\cite{rose} for ${\bf A}_{{\bf k},\mu}$ given by
Eq.~(\ref{kmu}), viz, 
\begin{equation}
{\bf A}_{{\bf k},\mu} = e^{i {\bf k} \cdot {\bf r}} 
{\hat {\pmb \varepsilon}}_{\mu}  = (2\pi)^{1/2} \sum_{L=1}^{\infty} 
\sum_{M=-L}^{L}(i)^{L} [L]
    D^{L}_{M \mu}(\phi,\theta,0) \Bigl[ {\bf A}^{(m)}_{LM}+ i \mu
\ {\bf A}^{(e)}_{LM} \Bigr]\ ,
\end{equation}
where $[L] =(2L+1)^{1/2}$ and $(\theta,\phi)$ denote polar angles of 
${\bf k}$, while ${\bf A}_{LM}^{(m)}$ and ${\bf A}_{LM}^{(e)}$
denote respectively the `magnetic' and `electric' $2^{L}$-pole 
solutions of the free Maxwell equations.
Using the notation
\begin{equation}
{\mathcal J}^{(m/e)}_{LM}(\omega)= \frac{Z\ e\ i^{L} [L]}
{2\pi {\sqrt {2\omega}}} \Bigl[ -i {\bf A}^{(m/e)}_{LM} \cdot 
{\pmb \bigtriangledown} + \frac{1}{2} {\pmb \sigma} \cdot 
\left({\pmb 
\bigtriangledown} \times 
{\bf A}^{(m/e)}_{{\bf k},\mu}\right) \Bigr]\ , \label{mathJ}
\end{equation}
and noting that Eq.~(\ref{mathJ}) is an irreducible tensor of rank 
$L$, 
we may apply the Wigner-Eckart theorem to write
\begin{eqnarray}
\langle J_{u}m_{u}|{\mathcal A}({\bf k},\mu)|J_{l}m_{l} \rangle=
{\mathcal A}({\bf}k,\mu)_{m_{u}m_{l}} 
& =& \sum_{L} C(J_{l}, L, J_{u};m_{l}, M, m_{u}) {\mathcal 
J}_{L}(\omega)
\nonumber \\ &&\times (i \mu)^{g_{+}(L)} 
    D^{L}_{M \mu}(\phi,\theta,0)\ ,
\label{absorption}
\end{eqnarray}
where the reduced matrix elements are given by
\begin{equation}
{\mathcal J}_{L}(\omega)= \langle J_{u}|| {\mathcal 
J}^{(m)}_{L}(\omega)||
J_{l} \rangle g_{-}(L) + \langle J_{u}|| {\mathcal 
J}^{(e)}_{L}(\omega)||
J_{l} \rangle g_{+}(L)\ , \label{rm}
\end{equation}
in terms of the projection operators
\begin{equation}
g_{\pm}(L)=\frac{1}{2}\Bigl[ 1\pm(-1)^{L} \pi_{u}\pi_{l} \Bigr]. 
\end{equation}
In the above equation $\pi_{u}, \pi_{l}$ denote the parities of the 
upper and 
lower levels. Using Eq.~(\ref{absorptiona}), we have
\begin{eqnarray}
\langle J_{l} m_{l}|{\mathcal E}({\bf k},\mu)|J_{u} m_{u} \rangle  =  
 {\mathcal E}({\bf k},\mu)_{m_{l}m_{u}} 
 &=& \sum_{L} C(J_{l}, L, J_{u}; m_{l}, M, m_{u}) {\mathcal J}_{L}
 (\omega)^{*}\nonumber \\&&\times(-i\mu)^{g_{+}(L)} 
     D^{L}_{M\mu}(\phi,\theta,0)^{*}\ . \label{emission}
\end{eqnarray}
Thus, we may express Eq.~(\ref{ephia}) as
\begin{equation}
T_{fi}({\bf k}\mu; {\bf k}' \mu') = \sum_{n} \phi_{n}
\sum_{m'_{f} m'_{i}} c^{s^{*}_{f}}_{m'_{f}} c^{s_{i}}_{m'_{i}}
\Bigl[ {\mathcal E}({\bf k},\mu) {\mathcal G}^{s_{n}}
{\mathcal A}({\bf k}',\mu') \Bigr]_{m'_{f}m'_{i}}\ , \label{Tmatrix}
\end{equation}
where the summation $\sum_{n}$ implies summation with respect to 
$s_{n}$ as well, and ${\mathcal A}({\bf k}',\mu')$
and ${\mathcal E}({\bf k},\mu)$ denote matrices, whose elements
\begin{eqnarray}
\langle J_{n}m'_{n}|{\mathcal A}({\bf k}',\mu')|J_{i}m'_{i} \rangle 
&=&
{\mathcal A}({\bf k}',\mu')_{m'_{n} m'_{i}}; \nonumber \\
& & \nonumber \\
\langle J_{f}m'_{f}|{\mathcal E}({\bf k},\mu)|J_{n}m''_{n} \rangle 
&=&
{\mathcal E}({\bf k},\mu)_{m'_{f} m''_{n}}\ , \label{elements}
\end{eqnarray}
may be written explicitly using Eqs.~(\ref{absorption}) and 
(\ref{emission}) and ${\mathcal G}$ denotes a hermitian $(2J_{n}+1)
\times (2J_{n}+1)$ matrix, which is defined in terms of its elements 
\begin{equation}
{\mathcal G}^{s_n}_{m''_n m'_n} = c^{s_{n}}_{m''_{n}}\ 
c^{s^{*}_{n}}_{m'_{n}}\ . \label{sn}
\end{equation}
Clearly, the summation over $n$ on right hand side of 
Eq.~(\ref{Tmatrix}) indicates a summation with respect to all the
atomic states $\{ \psi_{n} \}$, which constitute the complete 
orthogonal set. Since $n$ is a cumulative index
$\sum_{n}$ includes $\sum_{J_n}\ \sum_{s_n=1}^{2 J_n +1}$, apart 
from summation with respect to other quantum numbers. 
The left hand side of Eq.~(\ref{Tmatrix}) is written for a given 
$\psi_i$ and $\psi_f$ with energies $E_{i}$ and $E_{f}$ 
respectively. The 
quantities $s_{i}$ and $s_{f}$ are specified
by left hand side of Eq.~(\ref{Tmatrix}) and hence they are fixed 
entities on 
right hand side of Eq.~(\ref{elements}).

In the absence of the electric quadrupole field, the $s_{f}, s_{n}, 
s_{i}$ may be replaced respectively by
appropriate $M_{f}, M_{n}, M_{i}$ which are determined by the 
Kronecker $\delta$-function in Eq.~(\ref{special}), when
the magnetic field ${\bf B}$ alone is present and is directed along 
$(\theta_{B},\phi_{B})$. Thus,
$c^{s^{*}_f}_{m'_{f}}$ and $c^{s_i}_{m'_{i}}$ are replaced 
respectively by $D^{1}_{m'_f M_f}(\phi_{B},\theta_{B},0)^{*}$
and $D^{1}_{m'_i M_i}(\phi_{B},\theta_{B},0)$ with $M_{f}$ and 
$M_{i}$ being fixed by left hand side. It may be
noted that $\phi_{n}$ depends on $M_{n}$ and the summation over $n$ 
includes $\sum_{J_n}\ \sum_{M_n=-J_n}^{J_n}$,
with ${\mathcal G}^{s_n}$ replaced now by ${\mathcal G}^{M_n}$ whose 
elements are given by
\begin{equation}
{\mathcal G}^{M_n}_{m''_n m'_n}= D^{1}_{m''_{n} 
M_{n}}(\phi,\theta,0)\ 
D^{1}_{m'_{n} M_{n}}(\phi,\theta,0)^{*}\ . \label{dd}
\end{equation}

If the magnetic field ${\bf B}$ is along the $Z$-axis of ARF itself, 
the 
$s_{f}, s_{n}, s_{i}$ in Eq.~(\ref{Tmatrix})
may respectively be replaced by $m_{f}, m_{n}, m_{i}$ determined by 
the 
Kronecker $\delta$-function in 
Eq.~(\ref{simple}). Thus, $c^{s^*_f}_{m'_f}$ and 
$c^{s_i}_{m'_i}$ are replaced 
respectively by $\delta_{m'_f m_f}$ and 
$\delta_{m'_i m_i}$, where $m_f$ and $m_i$ are fixed by left hand 
side of 
Eq.~(\ref{Tmatrix}). Therefore, the summation with respect
to $m'_{f}$ and $m'_{i}$ on right hand side of Eq.~(\ref{Tmatrix}) 
drops after 
making the replacements $c^{s_f}_{m'_f}=1$ and
$c^{s_i}_{m'_i}=1$. The $\psi_{n}$ depends on $m_{n}$ and the 
summation 
over $n$ includes $\sum_{J_n}\ 
\sum^{J_n}_{m_n=-J_n}$, with ${\mathcal G}^{s_n}$ replaced by 
${\mathcal 
G}^{m_n}$, whose elements are
given by
\begin{equation}
{\mathcal G}^{m_n}_{m''_n m'_n}= \delta_{m''_n m_n} \ \delta_{m'_n 
m_n} \ ,
\label{GR}
\end{equation}
i.e., ${\mathcal G}^{s_n}$ gets replaced by a diagonal matrix with 
zeros  
everywhere except 
${\mathcal G}^{m_n}_{m_n m_n}=1$ in Eq.~(\ref{Tmatrix}).

\begin{figure}
 \begin{center}
 \includegraphics[width=14.2cm,height=9.cm]{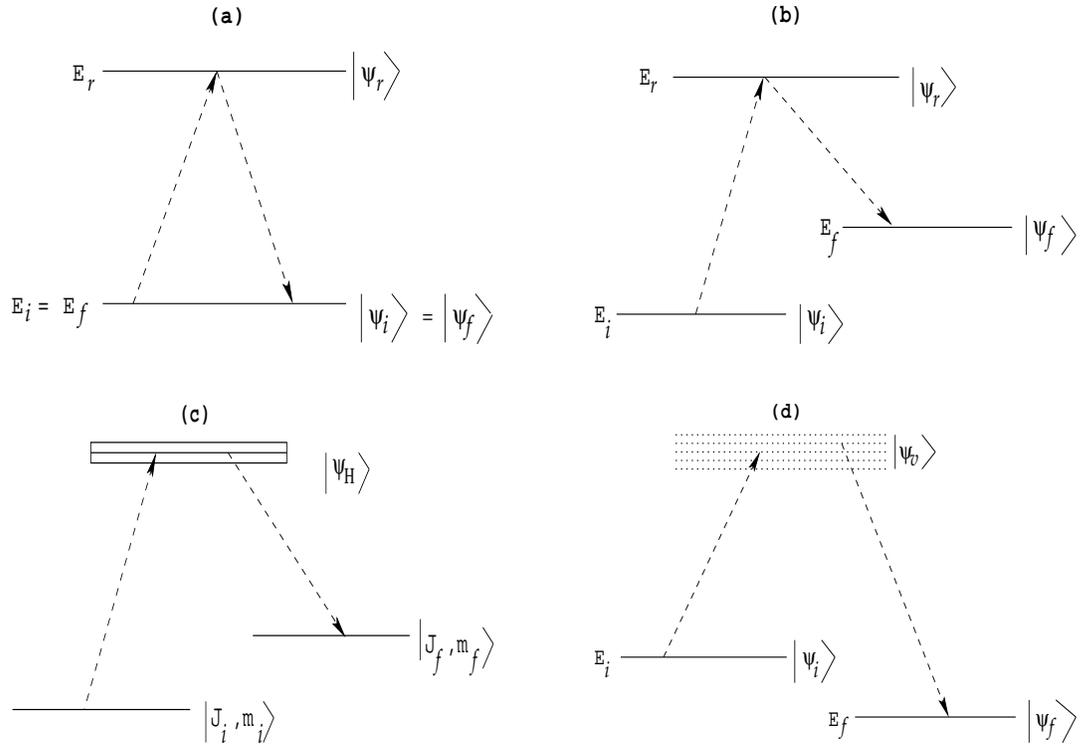}
 \end{center}
 \caption
{Level diagrams showing the atom-radiation interaction processes 
discussed in this 
paper. $(\bf{\rm a})$ two-level resonance scattering process,
$(\bf{\rm b})$ three-level fluorescence scattering process, 
$(\bf{\rm c})$ 
Hanle scattering 
process in weak magnetic fields and $(\bf{\rm d})$ the general case 
of 
Raman scattering.} 
 \label{diagram}
 \end{figure}

It may be noted that the atomic transitions from $\psi_{i}$ to 
$\psi_{n}$ 
following absorption of
$\omega'$ and from $\psi_{n}$ to $\psi_{f}$ consequent to the 
emission of 
$\omega$ are virtual transitions, 
which do not satisfy the celebrated Bohr criteria. This is in 
contrast to 
absorption or emission represented
by the $N=1$ term. They are real transitions which satisfy the Bohr 
criteria as already pointed out.
The summation over $n$ includes all atomic states $\psi_n$ with 
different 
energy eigenvalues $E_n$. However, all
of them do not contribute equally to Eq.~(\ref{Tmatrix}). 
The presence of $\phi_n$ on 
right hand side of Eq.~(\ref{Tmatrix}) indicates
that one has to pay more attention to contributions coming from 
those 
states $\psi_{n}$ with $E_{n}$ close to
$E_{i}+\omega'=E_{f}+\omega$. If there is an $E_n$ such that 
$E_{n}=E_{i}+\omega'=E_{f}+\omega$, the contribution
from this state alone overshadows all other contributions. The 
scattering 
is then referred to as resonance 
scattering. In particular, if $E_{i}=E_{f}$ the terminology `two 
level 
resonance scattering' is employed. This
is shown as $(a)$ in Fig.~\ref{diagram}, where $\omega'=\omega$. On 
the other 
hand if $E_{f}>E_{i}$ as in $(b)$ of Fig.~\ref{diagram},
the resonance scattering is referred to as three level resonance or 
fluorescence.

If there is no electric quadrupole field and the atom is exposed 
only to a 
pure magnetic field ${\bf B}$, such
that the $(2 J_{n}+1)$ states $|J_{n} M_{n} \rangle$ refer to 
distinctly 
separated energy levels as in
Zeeman effect, one can envisage resonance scattering taking place 
individually with each one of these
taking the role of the upper level, as shown in $(a)$ and $(b)$ of 
Fig.~\ref{diagram}, if the condition 
$E_{M_n}=E_{i}+\omega'=E_f+\omega$ is satisfied. On the other hand, 
if $gB 
< \Gamma_n$, the levels are not distinct
and all of them contribute coherently to form a single line. This is 
referred to as quantum interference in the
context of Hanle scattering, which is shown as $(c)$ in 
Fig.~\ref{diagram}. In 
contrast to Hanle effect where interference
occurs between magnetic substates with the same $J_{n}$, 
interference 
effects between states with different
$J_{n}$ have also been observed in polarization studies of Solar Ca 
II H-K 
and Na I D$_{1}$ and D$_{2}$ lines~\cite{stenflo3},
wherein it is mentioned that this can take place even when the lines 
are $3.5$ nm 
apart. The general terminology, `Raman
scattering' has been employed~\cite{stenflo2,stenflo3} to denote 
scattering, where 
contributions from several intermediate states
are involved. In general, therefore, we may rewrite 
Eq.~(\ref{ephia}) 
in the form 
\begin{equation}
T_{fi}({\bf k},\mu;{\bf k}',\mu')= \langle \psi_{f}|
{\mathcal E}({\bf k},\mu)|\psi_{v} \rangle \ ,
\end{equation}
where $|\psi_{v}\rangle$ represents a virtual state defined by
\begin{equation}
|\psi_{v} \rangle = \sum_{n} c^{v}_{n} |\psi_{n} \rangle ; 
\hspace{1cm}
|c^{v}_{n} \rangle = \phi_{n} \langle \psi_{n}|{\mathcal A}
({\bf k}',\mu')|\psi_{i}\rangle  \ ,
\end{equation}
which is clearly not an eigenstate of energy. In Raman effect, shown 
as 
$(d)$ in Fig.~\ref{diagram}, the lines corresponding to 
$E_{f}<E_{i}$ are 
referred to as anti-Stokes lines, in contrast to those with 
$E_{f}>E_{i}$ 
referred to as Stokes lines.  

\subsection{The dipole approximation}
If we neglect the spin-dependent second term in 
Eq.~(\ref{emissiona}) 
and employ dipole approximation 
$e^{i{\bf k}\cdot{\bf r}} \approx 1$ in ${\bf A}_{{\bf k}\mu}({\bf 
r})$
given by Eq.~(\ref{kmu}), then we may express 
Eq.~(\ref{absorptiona}) as 
\begin{equation}
\langle \psi_{u}|{\mathcal A}({\bf k},\mu)|\psi_{l} \rangle \approx
\langle \psi_{u}|{\hat {\pmb \varepsilon}}_{\mu} \cdot {\bf 
p}|\psi_{l} \rangle 
\approx \langle \psi_{l}|{\mathcal E}({\bf k},\mu)|\psi_{u} 
\rangle^{*}\ .
\end{equation}
In the above equation the momentum operator ${\bf p}=-i{\pmb 
\bigtriangledown}$ may 
be replaced by $[{\bf r},  H_{0}]$, to obtain 
\begin{equation}
\langle \psi_{u}|{\hat {\pmb \varepsilon}}_{\mu} \cdot [{\bf r},  
H_{0}]
|\psi_{l} \rangle \approx (E_l - E_u) \langle \psi_{u}|\psi_{l} 
\rangle ,
\end{equation}
if $E_u, E_l$ denote the energy eigenvalues of $\psi_u, \psi_l$ 
when considered as eigenstates of Eq.~(\ref{unperturbed}). 
We, thus, realize the
Kramers-Heisenberg form represented by Eq.~(\ref{unperturbed}) of 
\cite{stenflo4}.

\section{The scattering matrix for atoms in external electric 
quadrupole
and uniform magnetic fields}
The central result of the previous section is the derivation of the 
general expression for the 
on-energy-shell $T$-matrix
element $T_{fi}(\mu,\mu')$. If the incident radiation is in a pure
state
\begin{equation}
{\hat {\pmb \varepsilon}}_{i}= \sum_{\mu'} c^{i}_{\mu'} 
{\hat {\pmb \varepsilon}}'_{\mu'}\ ,
\end{equation}
the amplitude for detecting the scattered radiation in a pure state
\begin{equation}
{\hat {\pmb \varepsilon}}_{f}= \sum_{\mu} c^{f}_{\mu} 
{\hat {\pmb \varepsilon}}_{\mu}\ ,
\end{equation}
is given by 
\begin{equation}
T_{fi}({\hat {\pmb \varepsilon}}_{f},{\hat {\pmb \varepsilon}}_{i}) 
=\sum_{\mu \mu'}
c^{f^*}_{\mu}\  c^{i}_{\mu'}\ T_{fi}(\mu,\mu')\ ,
\end{equation}
where $\sum_{\mu'}|c^{i}_{\mu'}|^{2}=1=\sum_{\mu}|c^{f}_{\mu}|^{2}$. 
On the 
other hand, it is more convenient to employ the density matrix 
formalism~\cite{Mc,Oo1,Oo2}
to describe the states of polarization of the incident and
scattered radiation, as it is more general and can handle mixed 
states of 
polarization as well.

\subsection{The density matrix for polarized radiation}
The density matrix $\rho$ for polarized radiation may be written as
\begin{equation}
\rho=\frac{1}{2} \Bigl[ I + \sigma^{\gamma}_{x} Q + 
\sigma^{\gamma}_{y} U
+ \sigma^{\gamma}_{z} V \Bigr]= \frac{1}{2} \sum^{3}_{p=0}
 \sigma^{\gamma}_{p}\ S_{p}\ ,
\end{equation}
in terms of the well-known \cite{stenflo2} Stokes parameters 
$(I=S_{0}, Q=S_{1},U=S_{2}, V=S_{3})$ and Pauli matrices
$\sigma^{\gamma}_{x}=\sigma^{\gamma}_{1},
\sigma^{\gamma}_{y}=\sigma^{\gamma}_{2}, \sigma^{\gamma}_{z}=
\sigma^{\gamma}_{3}$ and 
the unit matrix $\sigma^{\gamma}_{0}$ whose rows and columns are 
labeled
by the left and right circular polarization states $|\mu=\pm 
1\rangle$
of radiation. Clearly,
\begin{equation}
S_{p}= tr (\sigma^{\gamma}_{p}\ \rho), \hspace{1cm} p=0,1,2,3 
\label{sp}
\end{equation}
where $tr$ denotes the trace or spur. A column vector ${\bf S}$
with elements $S_{p}, p=0,1,2,3$ is referred to as the Stokes vector 
for polarization. If we consider Eq.~(\ref{ephia}) as a $2 \times 2$ 
matrix $T$
with elements $T_{\mu \mu'} \equiv T_{fi}(\mu, \mu')$, the density
matrix $\rho$ of scattered radiation is given by
\begin{equation}
\rho=T\ \rho'\ T^{\dag} \ ,
\end{equation}
where $\rho'$ denotes the density matrix of polarized radiation 
incident on 
the atom. Using Eq.~(\ref{sp}), we have 
\begin{equation}
{\bf S}_{p}= \frac{1}{2} \sum^{3}_{p'=0} tr(\sigma^{\gamma}_{p}\ T 
\sigma^{\gamma}_{p'}\ T^{\dag}) {\bf S'}_{p'}\ , \label{tt}
\end{equation}
for the Stokes parameters of the scattered radiation, in terms of 
the
matrix $T$, its hermitian conjugate $T^{\dag}$ and the Stokes 
parameters
${\bf S'}_{p'}$ characterizing the radiation incident on the atom.

\subsection{The scattering matrix}
If the Stokes vector ${\bf S'}$ with elements $(I'=S'_{0}, 
Q'=S'_{1},
U'=S'_{2}, V'=S'_{3})$, characterizes the radiation incident on the
atom, the Stokes vector ${\bf S}$ characterizing the scattered 
radiation
may be expressed as 
\begin{equation}
{\bf S}= {\mathcal R}\ {\bf S'}\ , \label{srs}
\end{equation}
where the $4 \times 4$ matrix ${\mathcal R}$ is referred to as the 
scattering matrix.  Comparison of Eqs.~(\ref{tt}) and 
(\ref{srs}) readily identifies the elements of ${\mathcal R}$ as
\begin{equation}
{\mathcal R}_{pp'} = \frac{1}{2} \sum_{\mu \mu' \atop \mu'' \mu'''}
(\sigma^{\gamma}_{p})_{\mu'' \mu} T_{\mu \mu'} 
(\sigma^{\gamma}_{p})_{\mu' \mu'''} (T^{\dag})_{\mu''' \mu''}\ 
,\label{rpp}
\end{equation}
where we may use Eq.~(\ref{Tmatrix}) for $T_{\mu \mu'}$ and note 
that
$(T^{\dag})_{\mu''' \mu''}= T^{*}_{\mu'' \mu'''}$, for which we may
use the complex conjugate of Eq.~(\ref{Tmatrix}). We may thus write
\begin{eqnarray}
T_{\mu \mu'} &=& \sum_{n} \phi_{n} \sum_{m'_{f}m'_{i}} 
c^{s^{*}_{f}}_{m'_{f}} c^{s_{i}}_{m'_{i}} 
{\mathcal M}_{m'_{f}m'_{i}}(\mu, \mu')\ , \nonumber \\
(T^{\dag})_{\mu''' \mu''} &=& T^{*}_{\mu'' \mu'''} =
\sum_{n'} \phi^{*}_{n'} \sum_{m''_{f}m''_{i}} 
c^{s_{f}}_{m''_{f}} c^{s^{*}_{i}}_{m''_{i}} 
{\mathcal M}_{m''_{f}m''_{i}}(\mu'', \mu''')^{*}\ ,
\end{eqnarray}
where
\begin{eqnarray}
{\mathcal M}_{m'_{f}m'_{i}}(\mu, \mu') &=& \Bigl[ {\mathcal E}
({\bf k},\mu) {\mathcal G}^{s_{n}} {\mathcal A} ({\bf k}',\mu')
\Bigr]_{m'_{f} m'_{i}}\ , \nonumber \\
{\mathcal M}_{m''_{f} m''_{i}}(\mu'', \mu''')^{*} &=&
\Bigl[ {\mathcal E}
({\bf k},\mu'') {\mathcal G}^{s_{n'}} {\mathcal A} ({\bf k}',\mu''')
\Bigr]^{*}_{m''_{f} m''_{i}}\ , \nonumber \\
&=&  \Bigl[ {\mathcal A}^{\dag}
({\bf k}',\mu''') {\mathcal G}^{s_{n'}} {\mathcal E}^{\dag} 
({\bf k},\mu'')\Bigr]_{m''_{i} m''_{f}}\ ,
\end{eqnarray}
since ${\mathcal G}^{s_{n'}}$ is hermitian. Using the above in
Eq.~(\ref{rpp}), we have 
\begin{eqnarray}
\!\!\!\!\!\!{\mathcal R}_{pp'} &=& \frac{1}{2} \sum_{\mu \mu' \atop 
\mu'' \mu'''}
(\sigma^{\gamma}_{p})_{\mu'' \mu} (\sigma^{\gamma}_{p'})_{\mu' 
\mu'''}
\sum_{n n'} \phi_{n} \phi_{n'} 
    \sum_{m'_{f} m'_{i} \atop m''_{f} m''_{i}} \Bigl[ {\mathcal E}
({\bf k},\mu) {\mathcal G}^{s_{n}} {\mathcal A} ({\bf k}',\mu')
\Bigr]_{m'_{f} m'_{i}} \nonumber \\ &&\times c^{s_{i}}_{m'_{i}} 
c^{s^{*}_{i}}_{m''_{i}}
\Bigl[ {\mathcal A}^{\dag}
({\bf k}',\mu''') {\mathcal G}^{s_{n'}} {\mathcal E}^{\dag} 
({\bf k},\mu'')\Bigr]_{m''_{i} m''_{f}} 
c^{s_{f}}_{m''_{f}} c^{s^{*}_{f}}_{m'_{f}}\ . \label{rpp2}
\end{eqnarray}
Following Eq.~(\ref{sn}), we may define hermitian matrices 
${\mathcal G}^{s_{i}}$ and ${\mathcal G}^{s_{f}}$ through their 
elements
\begin{equation}
{\mathcal G}^{s_{i}}_{m'_{i} m''_{i}} =c^{s_{i}}_{m'_{i}}
c^{s^{*}_{i}}_{m''_{i}}\ , \label{cci}
\end{equation}
\begin{equation}
{\mathcal G}^{s_{f}}_{m''_{f} m'_{f}} = c^{s_{f}}_{m''_{f}}
c^{s^{*}_{f}}_{m'_{f}}\ , \label{ccf}
\end{equation}
so that we may rewrite Eq.~(\ref{rpp2}) as
\begin{eqnarray}
{\mathcal R}_{pp'} &=& \frac{1}{2} \sum_{\mu \mu' \atop 
\mu'' \mu'''} (\sigma^{\gamma}_{p})_{\mu'' \mu} 
(\sigma^{\gamma}_{p'})_{\mu' \mu'''} \sum_{n n'} \phi_{n} 
\phi^{*}_{n'}
\nonumber \\
& & Tr\Bigl[ {\mathcal E}({\bf k}, \mu) {\mathcal G}^{s_n}
{\mathcal A}({\bf k',\mu'}) {\mathcal G}^{s_{i}} {\mathcal A}^{\dag}
({\bf k',\mu'''}){\mathcal G}^{s_{n'}} {\mathcal E}^{\dag}({\bf k}, 
\mu'')
 {\mathcal G}^{s_f} \Bigr]\ , \label{final}
 \end{eqnarray}
where $Tr \equiv \sum_{m'_{f}}$ denotes the Trace or Spur of the 
$(2J_{f}+1) \times (2J_{f}+1)$ matrix within the square brackets, 
which
is defined through matrix multiplication of the eight matrices, each
of which is well-defined through Eqs.~(\ref{absorption}), 
(\ref{rm}),
(\ref{dd}), (\ref{cci}), (\ref{ccf}) for any specified 
atomic transition from an initial state
$\psi_{i}$ with energy $E_{i}$ and total angular momentum $J_{i}$ to
a final state $\psi_{f}$ with energy $E_{f}$ and total angular
momentum $J_{f}$, when the atom is exposed to a combined external
electric quadrupole field and a uniform magnetic field. It may be 
noted that $s_{i}$ and $s_{f}$ are fixed and the summation over 
$n,n'$
includes summation over $s_{n}, s_{n'}$.

\subsection{The particular case of resonance scattering via electric 
dipole
transitions between $J_{i}=J_{f}=0$ and $J_{n}=1$}
In this important particular case, which has often been investigated
in the presence of pure magnetic fields, it is clear that 
${\mathcal G}^{s_{i}}={\mathcal G}^{s_{f}}=1$ in Eq.~(\ref{final}) 
and 
$L=L'=1$ in Eqs.~(\ref{absorption}) and (\ref{rm}), so that we may 
write the trace appearing in Eq.~(\ref{final}) as
\begin{eqnarray}
& & Tr\Bigl[  {\mathcal A}({\bf k}',\mu') {\mathcal A}^{\dag}
({\bf k}',\mu''') {\mathcal G}^{s_{n'}} {\mathcal E}^{\dag}
({\bf k},\mu'') {\mathcal E}({\bf k},\mu) 
{\mathcal G}^{s_{n}} \Bigr] \nonumber \\&&= \sum_{m'_{n} m''_{n} 
\atop 
m'''_{n} m''''_{n}} 
           \Bigl[ {\mathcal A}({\bf k}',\mu') {\mathcal A}^{\dag} 
({\bf k}',\mu''') \Bigr]_{m'_{n} m'''_{n}} \Bigl({\mathcal 
G}^{s_{n'}}
\Bigr)_{m'''_{n} m''''_{n}} \nonumber \\
&&\times\Bigl[ {\mathcal E}^{\dag}({\bf k},\mu'') {\mathcal E} 
({\bf k},\mu) \Bigr]_{m''''_{n} m''_{n}}  
   \Bigl({\mathcal G}^{s_{n}} \Bigr)_{m''_{n} m'_{n}}\ .
\end{eqnarray}
We may use Eq.~(\ref{ni}), in combination with 
Eq.~(\ref{absorption}), to write
\begin{equation}
\Bigl[ {\mathcal A}({\bf k}',\mu') {\mathcal A}^{\dag} 
({\bf k}',\mu''') \Bigr]_{m'_{n} m'''_{n}} = 
|{\mathcal J}_{1}(\omega')|^{2} \mu' \mu''' D^{1}_{m'_{n} \mu'}
(\phi', \theta',0) D^{1}_{m'''_{n} \mu'''}(\phi', \theta',0)^{*} \ ,
\label{aadat}
\end{equation}
and Eq.~(\ref{fn}), in combination with Eq.~(\ref{emission}), to 
write
\begin{equation}
\Bigl[ {\mathcal E}^{\dag}({\bf k},\mu'') {\mathcal E} 
({\bf k},\mu) \Bigr]_{m''''_{n} m''_{n}} = 
|{\mathcal J}_{1}(\omega)|^{2} \mu \mu'' D^{1}_{m''''_{n} \mu''}
(\phi, \theta,0) D^{1}_{m''_{n} \mu}(\phi, \theta,0)^{*} \ . 
\label{eedat}
\end{equation}
Using Eq.~(\ref{sn}) for $\Bigl({\mathcal G}^{s_{n}}\Bigr)_{m''_{n} 
m'_{n}}$
and $\Bigl({\mathcal G}^{s_{n'}}\Bigr)_{m'''_{n} m''''_{n}}$, we may 
attach $c^{s^{*}_{n}}_{m'_{n}}c^{s_{n'}}_{m'''_{n}}$ along with 
$(\sigma^{\gamma}_{p'})_{\mu' \mu'''}$ to Eq.~(\ref{aadat}), while 
we may attach $c^{s_{n}}_{m''_{n}}c^{s^{*}_{n'}}_{m''''_{n}}$ along 
with
$(\sigma^{\gamma}_{p})_{\mu'' \mu}$ to Eq.~(\ref{eedat}), so that
\begin{eqnarray}
& & \sum_{\mu'\mu'''}(\sigma_{p'}^\gamma)_{\mu'\mu'''} \mu' \mu'''
\sum_{m'_{n} m'''_{n}} c^{s^{*}_{n}}_{m'_{n}} c^{s_{n'}}_{m'''_{n}}
D^{1}_{m'_{n} \mu'}(\phi',\theta',0)
D^{1}_{m'''_{n} \mu'''}(\phi',\theta',0)^{*} \nonumber \\
&&= \sum_{\lambda_{a}=0}^{2}
\sum_{\mu_{a}=-\lambda_{a}}^{\lambda_a} 
\sum_{m_{a}=-\lambda_{a}}^{\lambda_{a}} f_{p'}(\lambda_{a},\mu_{a})
F_{nn'}(\lambda_{a},m_{a})
D^{\lambda_{a}}_{m_{a}\mu_{a}}(\phi',\theta',0) \ ,
\end{eqnarray}
\begin {eqnarray}
& & \sum_{\mu\mu''}(\sigma_{p}^{\gamma})_{\mu''\mu}\mu''\mu
\sum_{m''_{n} 
m''''_{n}}c^{s_{n}}_{m''_{n}}c^{s^{*}_{n'}}_{m''''_{n}}
D^{1}_{m''''_{n} \mu''}(\phi,\theta,0)
D^{1}_{m''_{n} \mu}(\phi,\theta,0)^{*} \nonumber 
\\&&=\sum_{\lambda_{e}=0}^{2}
\sum_{\mu_{e}=-\lambda_{e}}^{\lambda_{e}} 
\sum_{m_{e}=-\lambda_{e}}^{\lambda_{e}} f_{p}(\lambda_{e},\mu_{e})
F_{n,n'}(\lambda_{e},m_{e})^{*}
D^{\lambda_{e}}_{m_{e}\mu_{e}}(\phi,\theta,0) \ .
\end{eqnarray}
Thus, we have 
\begin{eqnarray}
\label{scatmat-gen}
{\mathcal R}_{pp'}&=& \frac{1}{2}|{\mathcal J}_{1}(\omega)|^{2}
|{\mathcal J}_{1}(\omega')|^{2} \sum_{nn'} \phi_{n} \phi_{n'}
\sum_{\lambda_{a}=0}^{2} \sum_{\lambda_{e}=0}^{2}
F_{n,n'}(\lambda_{a},m_{a})F_{n,n'}(\lambda_{e},m_{e})^{*} \nonumber 
\\
& & f_{p}(\lambda_{e},\mu_{e}) f_{p'}(\lambda_{a},\mu_{a})
D^{\lambda_{e}}_{m_{e}\mu_{e}}(\phi,\theta,0)
D^{\lambda_{a}}_{m_{a}\mu_{a}}(\phi',\theta',0)\ ,
\end{eqnarray}
where 
\begin{equation}
F_{n,n'}(\lambda,m) = \sum_{m'_{n}} 
C(1,1,\lambda;m'_{n},-m'''_{n},m)
(-1)^{m'''_{n}}c^{s^{*}_{n}}_{m'_{n}} c^{s_{n'}}_{m'''_{n}}\ ,
\end{equation}
\begin{equation}
f_{p}(\lambda,\mu)=\sum_{\lambda} C(1,1,\lambda;\mu', -\mu''',\mu)
(-1)^{-\mu'''} \mu' \mu''' (\sigma^{\gamma}_{p})_{\mu' \mu'''}\ .
\end{equation}
The $f_{p}(\lambda,\mu)$ for $p=0,1,2,3$ may explicitly be written 
as 
\begin{eqnarray}
f_{0}(\lambda,\mu) &=& \frac{2}{\sqrt 3}\Bigl[ \delta_{\lambda,0} +
\frac{1}{\sqrt 2} \delta_{\lambda,2} \Bigr] \delta_{\mu,0}\ 
,\nonumber \\
f_{1}(\lambda,\mu) &=& - \delta_{\lambda,2} \Bigl[ \delta_{\mu,2} +
\delta_{\mu,-2} \Bigr]\ , \nonumber \\
f_{2}(\lambda,\mu) &=& i \delta_{\lambda,2} \Bigl[ \delta_{\mu,2}
- \delta_{\mu,-2} \Bigr]\ ,\nonumber \\
f_{3}(\lambda,\mu) &=& {\sqrt 2}\  \delta_{\lambda,1} 
\delta_{\mu,0}\ .
\end{eqnarray}
It is interesting to note that $F_{n,n'}(\lambda,m)$, in the 
particular
case of an atom exposed to a pure magnetic field ${\bf B}$ directed 
along 
$(\theta_{B},\phi_{B})$ may be written as
\begin{eqnarray}
\label{fnnp-mag}
F_{n,n'}(\lambda,m)&=&F_{M_{n},M'_{n}}(\lambda,m) \nonumber \\
&=& (-1)^{M'_{n}-m}
\ C(1,1,\lambda;M_{n},-M'_{n},M_{\lambda})\ 
\sqrt{4\pi}\ [\lambda]^{-1}\nonumber \\
& & Y_{\lambda M_{\lambda}}(\theta_{B},\phi_{B})\ ,
\end{eqnarray}
in terms of the spherical harmonics. From Eq.~(\ref{scatmat-gen}) we 
can 
recover the Hanle-Zeeman scattering matrix of Stenflo 
\cite{stenflo4} by setting 
the electric field strength to zero, and employing 
Eq.~(\ref{fnnp-mag}). 
The profile functions appearing in Eq.~(\ref{scatmat-gen}) are in 
the atomic 
frame. To obtain the Hanle-Zeeman scattering matrix, one needs to 
follow exactly 
the procedure outlined in \cite{stenflo4}, to transform $\phi_n$ to 
the 
laboratory frame through a Doppler convolution. In the weak field 
limit, we 
recover the well known Hanle scattering phase matrix of Landi 
Degl'Innocenti 
and Landi Degl'Innocenti \cite{lanandlan}, by assuming 
$\phi_{M_{n}}\phi^{*}_{M'_{n}}$ to be independent of $M_{n},M'_{n}$. 
In the 
strong field limit, $gB$ is large compared to the line widths. Hence 
the 
three Zeeman component lines are well separated. If we neglect the 
coupling 
between the Zeeman substates (drop the summation over $n$ in 
Eq.~(\ref{scatmat-gen}), by setting $n=n'$), we recover the 
restrictive phase 
matrix of Obridko~\cite{obridko}, which is basically a modified 
resonance 
scattering by individual Zeeman components.  

\section{Numerical Results and Discussions}
The calculations presented in this paper are applicable to magnetic 
fields of 
arbitrary strength, and also the presence of quadrupole electric 
fields 
surrounding the radiating atom. To check the correctness of our 
derivation, we 
have reproduced the results of Stenflo (1998, Fig.~3) for the 
particular case 
of Hanle-Zeeman effect \cite{ooth}. In weak magnetic fields, pure 
Hanle effect 
prevails. In strong fields, the Zeeman effect is the dominant 
process. In 
intermediate fields, there is a smooth transition from weak field 
Hanle effect 
to the strong field Zeeman effect. These two effects exhibit 
relative dominance 
in different regimes of field strength, but they fundamentally 
overlap over the 
entire regime.  

We consider the simplest case of a $J=0 \to 1 \to 0$ type 
transition which produces a standard Zeeman triplet. In this Section 
we 
present the results of a single scattering experiment  (see 
Eq.~\ref{srs}). 
We consider a $90^\circ$ scattering of an unpolarized 
beam of radiation incident on the atom. The incident radiation is 
also assumed 
to be frequency independent (broadband pulse). Since 
${\bf S}^\prime=(1,\, 0,\, 0,\, 0)^{\rm T}$, the scattered Stokes 
intensity 
components are nothing but $(I,\,Q,\,U,\,V)=({\mathcal R}_{11},\,
{\mathcal R}_{21},\, {\mathcal R}_{31},\, {\mathcal R}_{41})$, which 
measure 
the maximum degree of anisotropy for a given angle of scattering.     
The external magnetic field is assumed  
to be oriented along the $Z-$ axis of the astrophysical (laboratory)  
reference frame (see Fig.~\ref{geometry}). 

The scattering is assumed to be frequency coherent in the laboratory 
frame. A 
Voigt profile function with a damping parameter $a=0.004$ in units 
of the 
Doppler width $\Delta\nu_D=(\nu_0/c)\sqrt{2kT/M_a}$ is employed to 
compute all the results. In Sect.~4.1 we present 
the results for pure magnetic case. In Sect.~4.2 the results of 
scattering on atom immersed in pure electric quadrupole field is 
considered. 
Finally in Sect.~4.3 we consider the combined case of uniform 
magnetic and  
quadrupole electric fields. 

\subsection{The pure magnetic field case}
\begin{figure}
 \begin{center}
  \includegraphics[width=14cm,height=6.0cm]{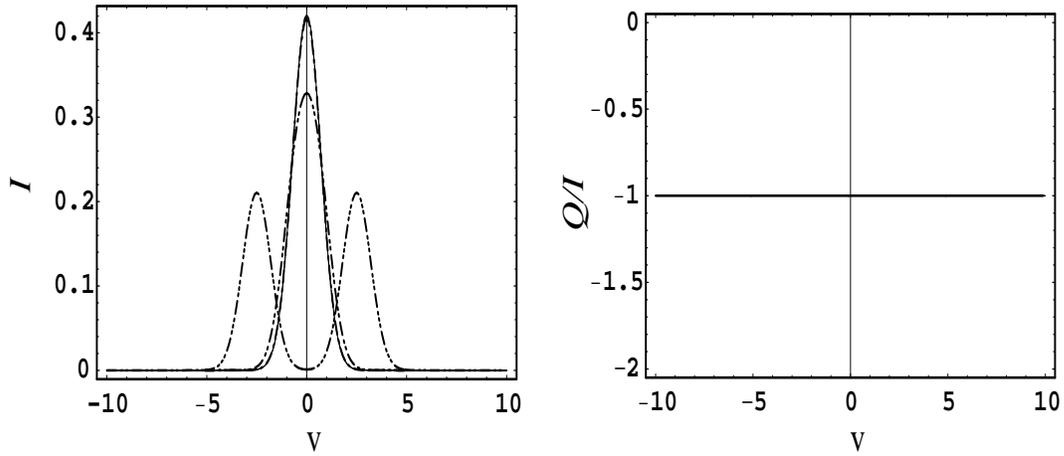}
 \end{center}
 \caption
{Stokes Profiles for Pure magnetic field case. Magnetic field is 
oriented along 
the $Z$ -axis of the ARF (see Fig.~\ref{geometry}). The model 
parameter 
employed are: $a = 0.004$, the scattering geometry defined by 
$\theta^\prime=0^\circ,\ \phi^\prime=0^\circ;\ \theta=90^\circ,\ 
\phi=45^\circ$, and the magnetic field 
strength defined by $v_B = 0.0008$ (solid line), 0.004 (dotted 
line), 0.02 
(dashed line), 0.1 (dash-dotted line), 
0.5 (dash-double dotted line), and, 2.5 (dash-triple dotted line). 
Notice a constant degree 
of linear polarization $Q/I = - 100\%$. }
 \label{puremag}
 \end{figure}

Fig.~\ref{puremag} shows singly scattered polarization profiles for 
this case. 
The field strength $B$ is chosen to represent the 
entire regime of Hanle-Zeeman effect ($v_B=0.0008$ -- $2.5$ in steps 
of a 
factor 5). The splitting parameter $v_B$ is defined as 
$v_B=\nu_L/\Delta\nu_D$ 
where $\nu_L=eB/4\pi mc$ is the Larmor frequency. 
The geometry of scattering chosen by us ($\theta^\prime=0^\circ,\ 
\phi^\prime=
0^\circ;\  \theta=90^\circ,\ \phi=45^\circ$) corresponds to Stokes 
$I\ {\rm and}\ Q$ given by 
\begin{equation}
I  =  {\frac {3}{8}} \big 
[\phi_1\phi^{*}_1+\phi_{-1}\phi^{*}_{-1}\big ]\,; \quad
Q  =  -I.  
\label{parti-caseq}
\end{equation}
The profile functions $\phi_1$ and $\phi_{-1}$, are now in the 
laboratory 
frame. They are obtained by a convolution of the atomic frame 
Lorentzian (see 
Eq.~(\ref{prof-fun})) with the Doppler profile. The real part of 
$\phi_{1}$ for example, is a Voigt function whereas the imaginary 
part is a 
Faraday-Voigt function (see \cite{stenflo4}). In the laboratory 
frame 
the frequency is expressed in dimensionless units 
(${\rm v}=(\nu_0-\nu)/\Delta\nu_D$). Clearly, for weak fields the 
Zeeman 
splitting is not complete. Hence the $I$ profile simply broadens 
without 
exhibiting a separation of the components. This kind of intensity 
profiles are 
very typical of Hanle and Hanle-Zeeman regime in the second solar 
spectrum of 
the Sun (see \cite{stenflo2}). The case of $v_B = 2.5$ represents a 
strong 
field Zeeman effect, and we clearly see a well separated doublet. 
Since the line of sight (the scattered ray) is perpendicular to the 
magnetic 
field, according to conventional Zeeman effect theory (Zeeman effect 
treated as 
absorption/emission), one expects a triplet pattern in $I$ profile, 
and a $Q$ 
profile with $\pi$ component having opposite polarization compared 
to the two 
$\sigma$ components, along with $U=V=0$. However, we now observe 
only a 
doublet in the $I$ profile (see Fig.~\ref{puremag}), showing that 
the 
mechanism involved is indeed a `scattering' process and not 
`absorption 
followed by uncorrelated emission' process. With the help of 
classical theory 
of dipole scattering, one can argue that, for $90^\circ$ scattering 
and for a 
magnetic field along the $Z$ - axis, the $\pi$ component is not 
excited at all 
by the incident radiation. Only the components with electric 
vibration 
perpendicular to the scattering plane (containing the incident and 
scattered 
ray) are excited, and hence the two $\sigma$ components 
appear in the $I$ profile. 

From Eq.~(\ref{parti-caseq}) and also from the Fig.~\ref{puremag}, 
we see that 
$Q/I = -1$, i.e., independent of frequency as well as the field 
strength. In 
other words scattered polarization is same as the well known 
non-magnetic pure 
Rayleigh scattering polarization. This is to be expected, because 
in the weak field regime, the Hanle effect is absent for vertical 
magnetic 
fields, and the scattered polarization can arise only due to 
Rayleigh 
scattering process. In the strong field regime, both the sigma 
components have 
linear polarization of equal magnitude, and scatter independently 
(see 
Eq.~(\ref{parti-caseq})). Therefore we obtain a maximum degree of 
linear 
polarization, namely $Q/I=-1$. We refer to this case as `Zeeman 
scattering' or 
equivalently  `Rayleigh scattering in strong magnetic fields' (see 
also 
\cite{bom-2,sam1}). 

\subsection{The case of pure electric quadrupole field}
\begin{figure}
 \begin{center}
   \includegraphics[width=14cm,height=10.cm]{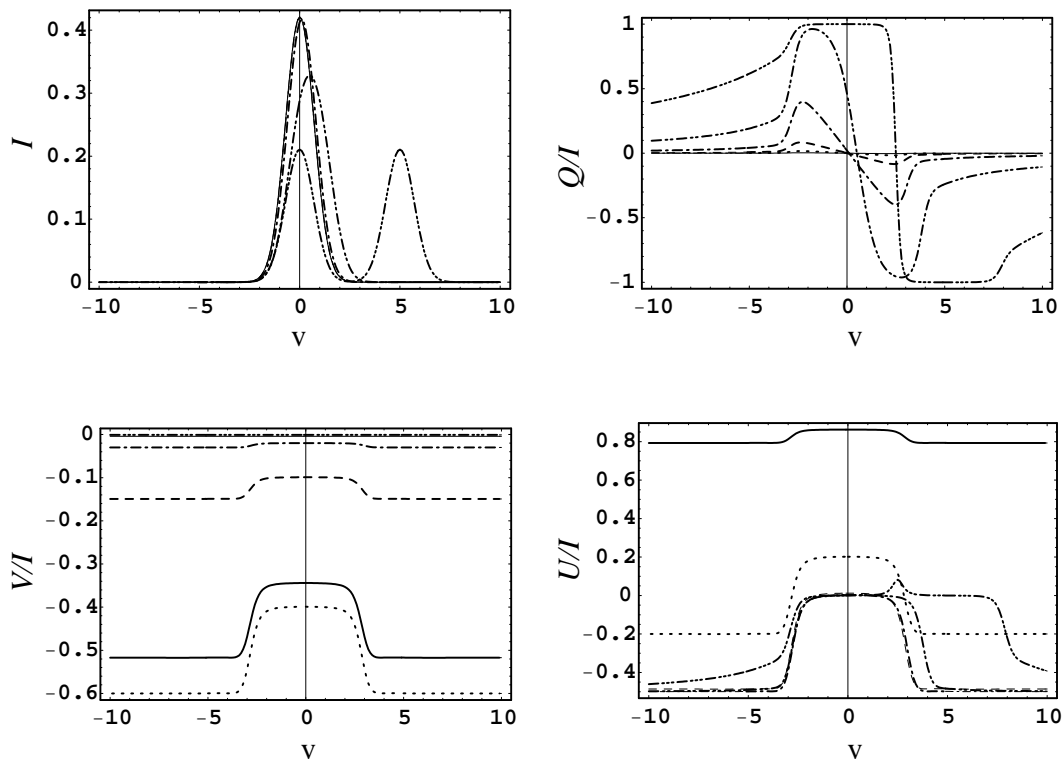}
 \end{center}
 \caption
{Stokes line profiles for pure quadrupolar electric field case. 
The model parameters employed are: $a = 0.004$, with the scattering 
geometry 
$\theta^\prime=0^\circ,\ \phi^\prime=0^\circ;\ \theta=90^\circ,\ 
\phi=45^\circ$, and the electric field 
strength defined through the splitting parameter $v_A = 0.0008$ 
(solid line), 
0.004 (dotted line), 0.02 (dashed line), 0.1 (dash-dotted line), 0.5 
(dash-double dotted line), and, 2.5 (dash-triple dotted line), and 
the 
asymmetry parameter $\eta = 1$. Notice the blue shift of the 
profiles compared to the pure 
magnetic field case for large values of $v_A$ ($>0.1$). }
 \label{pureelec}
 \end{figure}

Fig.~\ref{pureelec} shows the singly scattered Stokes profiles in 
the 
presence of a pure electric quadrupole field surrounding the atom. 
The single 
scattering experiment is considered as in the pure magnetic field 
case. The 
electric field is characterized by two parameters, namely the 
electric 
splitting parameter $A$ which is taken in the same way as magnetic 
splitting 
parameter, namely $v_A=0.0008$ -- 2.5 in steps of 5, where $v_A$ is 
electric splitting parameter in Doppler width units. The asymmetry 
parameter 
$\eta$ is fixed at 1. The upper $J_n = 1$ level splits into 3 levels 
with 
energies $-2A,\ 0$ and $+2A$ corresponding to the eigenstates 
$\psi_1=|1,0\rangle$, $\psi_2=(|1,-1\rangle-|1,1\rangle)/\sqrt{2}$, 
and 
$\psi_3=(|1,-1\rangle+|1,1\rangle)/\sqrt{2}$, 
respectively (see Fig.~2c of \cite{Oo1}). For the scattering 
geometry 
employed by us, the Stokes parameters are given by the analytic 
expressions 
\begin{eqnarray}
\label{parti-elec}
I = {\frac{3}{8}} \big[\phi_2\phi^{*}_2 + \phi_3\phi^{*}_3 \big 
];&\quad &
Q = {\frac{3}{8}} \big[\phi_2\phi^{*}_2 - \phi_3\phi^{*}_3 \big 
],\nonumber \\
U = {\frac{3}{8}} \big[\phi_2\phi^{*}_3 + \phi_3\phi^{*}_2 \big 
];&\quad &
V = {\frac{3}{8}} i\big[\phi_2\phi^{*}_3 - \phi_3\phi^{*}_2 \big ].
\end{eqnarray} 
The profile functions $\phi_2$ and $\phi_3$ correspond to the 
eigenstates 
$\psi_2$ and $\psi_3$ respectively, and are given by 
\begin{equation}
\label{prof-elec}
\phi_{2,3} = H({\rm v}+(1\mp\eta)v_A , a) + 2\, i\,  F({\rm 
v}+(1\mp\eta)v_A , a),
\end{equation}
in the laboratory frame. 
Here $H$ and $F$ are the well known Voigt and Faraday-Voigt 
functions 
\cite{stenflo4}. From Eq.~(\ref{parti-elec}), we note that $U$ and 
$V$ are 
generated purely due to the coupling between the eigenstates 
$\psi_2$ and 
$\psi_3$.

For weak electric fields ($v_A < 0.1$) the shapes of the $I$ profile 
are not 
affected significantly, when compared to the corresponding pure 
magnetic case 
(see Fig.~\ref{puremag}). For $v_A > 0.1$ there is a blue shift 
compared to 
the pure magnetic case, which increases gradually for larger values 
of $v_A$ 
until we get a doublet that is asymmetrically placed 
about the line center. The $Q/I$ profiles have an interesting shape, 
in the 
sense that they are similar to the $V/I$ profiles of the pure 
magnetic field 
case. For $v_A=0.0008$ and 0.004, $Q/I$ is extremely small. 
As $v_A$ increases, $Q/I$ gradually increases, and the zero 
cross-over point 
of $Q/I$ shifts toward the blue (see for eg. dash-triple dotted line 
in Fig.~\ref{pureelec}). $U/I$ and $V/I$ show interesting behavior. 
Unlike $Q/I$, the ratio $U/I$ takes largest value for $v_A=0.0008$, 
and is 
entirely positive (see upper most solid line in $U/I$ panel of 
Fig.~\ref{pureelec}). For $v_A=0.004$ (dotted line) $U/I$ takes both 
positive 
and negative values. As $v_A$ increases we note that $U/I$ goes to 
zero in the 
line core (except for $v_A=2.5$ -- dash-triple dotted line which 
shows a 
small positive peak around ${\rm v}=2.5$), and becomes entirely 
negative and nearly constant 
in the wings. The shape of $V/I$ is similar to that of $U/I$ 
(namely, taking 
constant values at line core and line wings with a swift transition 
around 
${\rm v}\simeq 3$). The magnitude of $V/I$ initially increases with 
$v_A$ and 
then rapidly decreases toward zero, as $v_A$ increases. For 
$v_A>0.1$, the 
$V/I$ is nearly zero throughout the line profile (see 
Fig.~\ref{pureelec}). 

If the 
mechanism involved is a pure emission process, then one would expect 
all the 
three wave functions $\psi_{1,2,3}$ to contribute, and produce line 
components 
in Stokes $I$, along with $Q=V=0$, and $U\ne 0$ (see Eqs.~(69)--(72) 
in 
\cite{Oo1}). However when the interaction of 
radiation is treated as scattering (represented through angular 
correlations 
between incident and scattered ray), we see that 
only a doublet is seen in Stokes $I$ (since $\psi_1$ is not excited 
at all 
according to Eq.~(\ref{parti-elec})) with $Q,\ U$ and $V$ non-zero. 
This is the essential difference between the spontaneous emission 
process and 
the scattering process in quadrupole electric fields. We have also 
computed 
the Stokes profiles for the $\eta=0.5$ case, and find that they do 
not differ 
qualitatively from $\eta=1$ case, except for changes caused by 
different 
amount of level splitting. 

\subsection{The case of combined magnetic and quadrupole electric 
fields}
\begin{figure}
 \begin{center}
  \includegraphics[width=14cm,height=10.0cm]{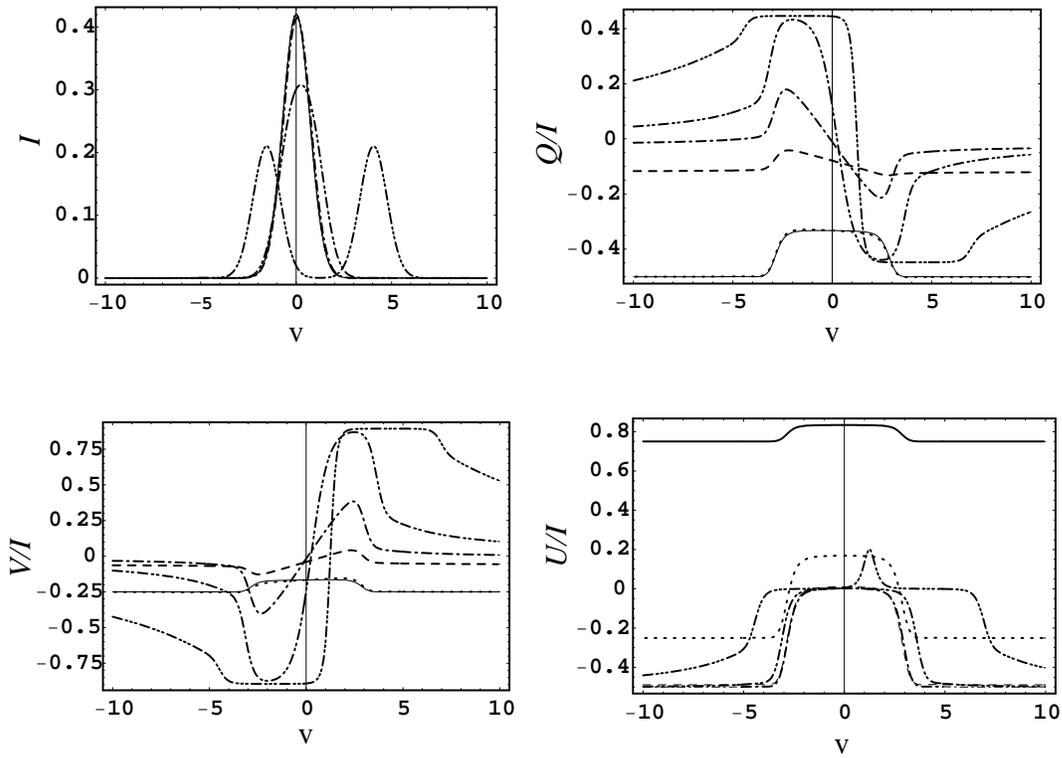}
 \end{center}
 \caption
{Stokes Profiles for combined quadrupolar electric and uniform 
magnetic field 
case. The ratio $r = A/B = 0.5$ and the asymmetry parameter $\eta = 
1$. 
Other input model parameters, scattering geometry and the 
line types are same as in Fig.~\ref{puremag}. }
 \label{magelec}
 \end{figure}

\begin{figure}
 \begin{center}
  \includegraphics[width=8cm,height=14.0cm]{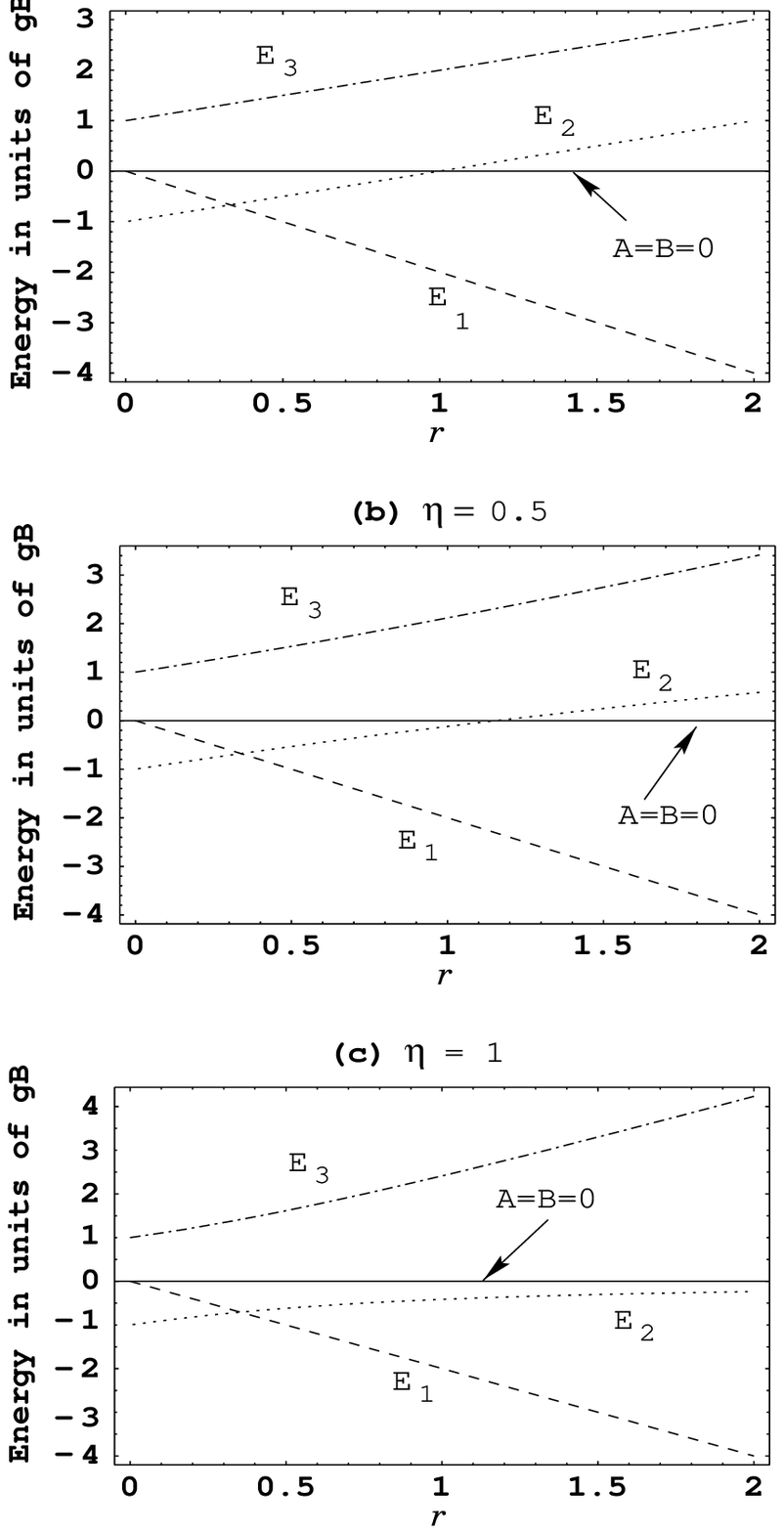}
 \end{center}
 \caption
{Energy level diagrams showing the combined effect of magnetic and 
electric 
quadrupole fields. The panels (a, b, c) represent different 
asymmetry 
parameters $\eta$. The energies $E_1$, $E_2$, and $E_3$ 
corresponding to the 
states $\psi_1$, $\psi_2$, and $\psi_3$ respectively are plotted. 
The 
value $r=0$ corresponds to the pure Zeeman case. }
 \label{crossing}
 \end{figure}

In Fig.~\ref{magelec} we show the Stokes profiles for this case. We 
employ 
ratio $A/B=0.5$ (which defines the relative strength of electric 
field with 
respect to the magnetic field), and the asymmetry parameter 
$\eta=1$. The 
splitting parameter $v_B$ ($= 0.0008$ -- 2.5) is employed as in the 
pure 
magnetic case. The simplest geometry of the combined magnetic and 
quadrupole 
fields with ${\bf B}$ along the Z-axis of the PAF, and PAF itself 
coinciding 
with the ARF, is employed in the computation of the results in this 
section. 
The scattering geometry is same as in Figs.~\ref{puremag} and 
\ref{pureelec}. 
The upper level $J_u = 1$ is split into three levels $n=1,2,3$ with 
energies 
\begin{equation}
E_1 = -2rB\,; \quad\quad E_{2,3}=[r \mp (r^2\eta^2+1)^{1/2}]B,  
\end{equation}
where $r = A/B$. The corresponding eigenstates are given by (see 
\cite{Oo1}) 
\begin{equation}
\psi_1 = |1,0\rangle\,; \quad \psi_2 = b_1|1,-1\rangle + 
b_2|1,1\rangle\,; 
\quad \psi_3 = -b_2|1,-1\rangle + b_1|1,1\rangle, 
\end{equation}
where the interference coefficients are defined by 
\begin{eqnarray}
b_1 = {\frac{r\eta + 1 + (r^2\eta^2+1)^{1/2}} 
    {2\bigg[r^2\eta^2+1+r\eta(r^2\eta^2+1)^{1/2}\bigg]^{1/2}}},  
\nonumber\\
b_2 = {\frac{r\eta - 1 + (r^2\eta^2+1)^{1/2}} 
    {2\bigg[r^2\eta^2+1+r\eta(r^2\eta^2+1)^{1/2}\bigg]^{1/2}}}. 
\end{eqnarray}
Fig.~\ref{crossing} shows the energy level splitting for spin-1 
upper 
level, exposed to the simultaneous presence of an external electric 
quadrupole 
and uniform magnetic fields. Notice that ``level-crossing" occurs 
when 
the electric field strength increases. It is worth noting that these 
levels 
are not pure states, but superposed states. The cross-over occurs 
for only a 
single value of electric quadrupole field strength at which the 
levels 
become degenerate. 

Fig.~\ref{magelec} presented in this paper can be 
interpreted using the panel (c) of Fig.~\ref{crossing}. For $r=0.5$, 
the 
energy eigenstates $E_1$ and $E_2$ are below the $A=B=0$ reference 
line. For 
the geometry chosen, the state $\psi_1=|1,0\rangle$ is not 
excited, as can be seen from the following analytic expressions for 
the scattered Stokes 
parameters\,:
\begin{eqnarray}
\label{parti-elecmag}
I&=&{\frac{3}{8}} \big[\phi_2\phi^{*}_2 + \phi_3\phi^{*}_3 \big 
],\nonumber\\
Q&=&{\frac{3}{8}} \big[ 2\, b_1\, b_2\, (\phi_3 
\phi^{*}_3-\phi_2\phi^{*}_2) + 
i\, (b_1^2-b_2^2) (\phi_3\phi^{*}_2-\phi_2\phi^{*}_3 )\big 
],\nonumber \\
U&=&{\frac{3}{8}} \big[\phi_2\phi^{*}_3 + \phi_3\phi^{*}_2 \big 
],\nonumber\\
V&=&{\frac{3}{8}} \big[(b_1^2-b_2^2)(\phi_2\phi^{*}_2-\phi_3 
\phi^{*}_3)+2\,i \,
b_1\, b_2\,(\phi_3\phi^{*}_2 - \phi_2\phi^{*}_3) \big ].
\end{eqnarray}
The profile functions $\phi_2$ and $\phi_3$ in the above expressions 
correspond to the eigenstates $\psi_2$ and $\psi_3$ respectively, 
and 
are given by 
\begin{equation}
\label{prof-elecmag}
\phi_{2,3} = H\bigg({\rm v}+\big(r\mp\sqrt{r^2\eta^2+1}\big)v_B , 
a\bigg) + 2\,i \,
F\bigg({\rm v}+\big(r\mp\sqrt{r^2\eta^2+1}\big)v_B , a\bigg),
\end{equation}
in the laboratory frame for any $\eta$.  

The Stokes $I$ profiles in the combined case are quite similar in 
amplitude 
and shape to those in the pure magnetic case, except for the 
position of the 
component lines (for $v_B>0.1$). For $v_B=0.0008$ and 0.004 (solid 
and dotted 
lines which nearly overlap on each other in $Q/I$ panel 
of Fig.~\ref{magelec}), the $Q/I$ profiles are very different 
compared to the 
corresponding pure electric field case, and are entirely negative. 
For 
$v_B>0.004$, these $Q/I$ profiles resemble their corresponding 
counterparts of 
the pure electric field case in terms of the frequency dependence. 
Clearly, 
there is a decrease in the magnitude of $Q/I$ when compared to both 
the pure 
magnetic and the pure electric case. 

$U/I$ in the combined case is quite similar to the pure electric 
case (compare 
$U/I$ in Eq.~(\ref{parti-elec}) with Eq.~(\ref{parti-elecmag}) for 
$r=0.5$ 
and $\eta=1$). In 
other words, we can conclude that for this particular geometry the 
entire 
frequency dependence of the $U/I$ comes from the electric field 
effect. The 
only difference is in $v_B=2.5$ (dash-triple dotted line) case, 
where in the small peak 
is enhanced as well as it is now centered around 1.5, unlike in pure 
electric 
field case. $V/I$ profiles in combined case bear resemblance to 
$Q/I$ 
profiles in shape, differing only in magnitude and sign (for $v_B > 
0.0004$). 
Clearly, the quadrupolar electric field produces an additional 
depolarization 
in $Q/I$ (compared to the $Q/I$ of pure magnetic case in 
Fig.~\ref{puremag}), 
and a rotation of the plane of polarization (or generation of new 
$U/I$). 
Therefore in the special geometry of vertical magnetic fields (${\bf 
B}$ 
parallel to the Z-axis of PAF and ARF simultaneously), if one 
observes 
strong $U/I$ and $V/I$ signals as well as relatively smaller $Q/I$ 
signals in a 
$90^\circ$ scattering (eg. as in the case of extreme limb 
observations of the 
Solar chromosphere; or like the special scattering geometry that we 
considered 
for discussion), it could indicate 
the asymmetries arising from the quadrupolar electric fields 
surrounding the 
atom. Notice that in the absence of electric fields, $U/I=V/I=0$ and 
$Q/I=-1$ 
for the vertical magnetic fields (see Fig.~\ref{puremag}). Another 
diagnostic 
indicator of quadrupolar electric fields is the net blue shift of 
the Stokes 
profiles, unlike the linear Stark effect which produces symmetric 
shifts with 
respect to the line center. 

For the sake of discussion, we have computed the scattered Stokes 
profiles for 
$\eta=0$ also in the combined case. This case is interesting, 
because, from 
Fig.~\ref{crossing} we observe that for $r\le 0.5$, the splitting is 
independent of $\eta$, namely the splitting pattern is same for both 
$\eta=0$ 
and 1. As a result, the Stokes $I$ for both $\eta = 0$ and 1 cases 
are nearly 
identical. This can be understood from Eq.~(\ref{parti-elecmag}) by 
setting $\eta=0$. 
The profile functions $\phi_{2,3}$ are now given by 
Eq.~(\ref{prof-elecmag}) with $\eta=0$. We note that the real 
part of $\phi_{2}$ for $\eta=0$ and 1 respectively are 
$H({\rm v}-0.5v_B,a)$ and $H({\rm v}-0.62v_B,a)$. Similarly for 
$\phi_3$, 
corresponding real parts are $H({\rm v}+1.5v_B,a)$ and $H({\rm 
v}+1.62v_B,a)$. 
For this reason, the Stokes $I$ as well as $U/I$ profile for 
$\eta=0$ and 1 are nearly identical. 
However the $Q/I$ and $V/I$ profiles  
are quite different (see Eqs.~(\ref{parti-elecmag})).   
The $Q/I$ profiles of $\eta = 0$ case have shapes and 
magnitudes similar to the $V/I$ profiles of Fig.~\ref{pureelec}. 
The $V/I$ profiles of $\eta=0$ case have shapes and 
magnitudes similar to $Q/I$ profiles of Fig.~\ref{pureelec}, 
except for a sign difference. 

\section{Conclusions}
The scattering matrices for the combined effect of electric 
quadrupole field and 
uniform magnetic fields (of arbitrary strength) are derived using 
quantum 
electrodynamic approach. The scattering matrix for Hanle-Zeeman 
effect is 
validated by comparing with the published results of Stenflo 
\cite{stenflo4}. The 
quadrupole electric field is characterized by strength and asymmetry 
parameters, which produce unique diagnostic signatures that may be 
employed to 
detect the electric charge distribution asymmetries in the solar 
atmosphere. 
The theoretical formulation is quite general and can handle not only 
the 
simplest case of a triplet ($J=0\to 1 \to 0$) transition that is 
employed for 
illustrations in this paper, but also an arbitrary choice of quantum 
numbers. 
We have demonstrated the properties of the coherence or interference 
phenomena, like strong field Zeeman scattering, Hanle and magnetic 
Raman 
scattering that are important in the interpretation of spectral 
lines in the 
second solar spectrum. 

\ack GR is grateful to Professors~B.~V. Sreekantan, R. Cowsik, J.~H. 
Sastry, 
R. Srinivasan and S.~S. Hasan for facilities provided for research 
at the 
Indian Institute of Astrophysics. Yee Yee Oo wishes to thank the 
Director of 
Indian Institute of Astrophysics, the Chairman, Physics Dept., 
Bangalore 
University for extending the research facilities. She also 
acknowledges ICCR, 
Government of India for financial support to visit India, during 
which the 
major part of the work was carried out. MS is financially supported 
by 
Council of Scientific and Industrial Research (CSIR), through a SRF 
(Grant 
No\,:9/890(01)/2004-EMR-I), which is gratefully acknowledged. 
Authors would like to thank Mr. K. Nagaraju for useful discussions.


\newpage


\begin{thebibliography}{}
\bibitem{hale} Hale GE. On the probable existence of a magnetic 
field in 
sun-spots.  Astrophys J 1908;28;315-43. 
\bibitem{chandra} Chandrasekhar S. Radiative Transfer. Oxford 
Clarendon Press; 1950.   
\bibitem{landi-1} Landi Degl'Innocenti E. Polarization in spectral 
lines. 
I - A unifying theoretical approach. Sol Phys 1983a;85:3-31. 
\bibitem{landi-2} Landi Degl'Innocenti E. Polarization in Spectral 
Lines. 
II: A Classification Scheme for Solar Observations. Sol Phys 
1983b;85:33-40. 
\bibitem{landi-3} Landi Degl'Innocenti E. Polarization in spectral 
lines. 
III - Resonance polarization in the non-magnetic, collisionless 
regime. 
Sol Phys 1984;91;1-26. 
\bibitem{landi-4} Landi Degl'Innocenti E. Polarization in spectral 
lines. 
IV - Resonance polarization in the Hanle effect, collisionless 
regime. 
Sol Phys 1985;102:1-20. 
\bibitem{lbs1} Landi Degl'Innocenti E, Bommier V, 
Sahal-Br$\acute{\rm e}$shot S. Resonance line polarization and the 
Hanle 
effect in optically thick media. I - Formulation for the two-level 
atom. 
Astron Astrophys 1990;235:459-71. 
\bibitem{lbs2} Landi Degl'Innocenti E, Bommier V, 
Sahal-Br$\acute{e}$chot S. Resonance line polarization for arbitrary 
magnetic 
fields in optically thick media. I - Basic formalism for a 
3-dimensional 
medium. Astron Astrophys 1991a;244:391-400. 
\bibitem{lbs3} Landi Degl'Innocenti E, Bommier V, 
Sahal-Br$\acute{e}$chot S. Resonance Line Polarization for Arbitrary 
Magnetic Fields in Optically Thick Media - Part Two - Case of a 
Plane-Parallel 
Atmosphere and Absence of Zeeman Coherences. 
Astron Astrophys 1991b;244:401-08. 
\bibitem{bom-1} Bommier V. Master equation theory applied to the 
redistribution of polarized radiation, in the weak radiation field 
limit. 
I. Zero magnetic field case. Astron Astrophys 1997a;328:706-25.
\bibitem{bom-2} Bommier V. Master equation theory applied to the
redistribution of polarized radiation, in the weak radiation field 
limit.
II. Arbitrary magnetic field case. Astron Astrophys 
1997b;328:726-51. 
\bibitem{stenflo2} Stenflo JO. Solar Magnetic Fields- Polarized 
Radiation Diagnostics. Dordrecht: Kluwer Academic Publishers; 1994.
\bibitem{stenflo3} Stenflo JO. Quantum interferences, hyperfine 
structure, and 
Raman scattering on the Sun. Astron Astrophys 1997;324:344-56. 
\bibitem{stenflo4} Stenflo JO. Hanle-Zeeman scattering matrix. 
Astron Astrophys 
1998;338;301-10. 
\bibitem{bs99} Bommier V, Stenflo JO. Partial frequency 
redistribution with 
Hanle and Zeeman effects. Non-perturbative classical theory. 
Astron Astrophys 1997;350:327-33. 
\bibitem{favati} Favati B, Landi Degl'Innocenti E, Landolfi M. 
Resonance 
scattering of Lyman-alpha in the presence of an electrostatic field. 
Astron Astrophys 1987;179:329-38. 
\bibitem{casini} Casini R, Landi Degl'Innocenti E. The polarized 
spectrum of 
hydrogen in the presence of electric and magnetic fields. Astron 
Astrophys 
1993;276:289-302.
\bibitem{casini1} Casini R. Resonance scattering formalism for the 
hydrogen 
lines in the presence of magnetic and electric fields. Phys Rev A 
2005;71:062505-1-17.
\bibitem{brillant} Brillant S, Mathys G, Stehle C. Hydrogen line 
formation in 
dense magnetized plasmas. Astron Astrophys 
1998;339:286-97. 
\bibitem{tru} Trujillo Bueno J, Moreno Insertis F, S$\acute{\rm 
a}$nchez F.
editors. Astrophysical Spectropolarimetry. Cambridge: Cambridge 
University 
Press, 2002.
\bibitem{ll} Landi Degl'Innocenti E, Landolfi M. Polarization in
Spectral Lines. Dordrecht: Kluwer Academic Publishers, 2004.
\bibitem{sam1a} Sampoorna M, Nagendra KN, Stenflo JO. Hanle--Zeeman 
Redistribution Matrix I. Classical theory expressions in the 
laboratory frame. 
Astrophys J 2006 (submitted).
\bibitem{sam1} Sampoorna M, Nagendra KN, Stenflo JO. Hanle--Zeeman 
Redistribution Matrix II. Comparison of Classical and Quantum 
Electrodynamic 
Approaches. Astrophys J 2006 (submitted).
\bibitem{Oo1} Yee Yee Oo, Nagendra KN, Sharath Ananthamurthy,
Vijayashankar R, Ramachandran G. Polarization of line radiation in 
the
presence of external electric quadrupole and uniform magnetic 
fields. JQSRT
2004;1:35-64.
\bibitem{Oo2} Yee Yee Oo, Nagendra KN, Sharath Ananthamurthy, 
Swarnamala
Sirsi, Vijayashankar R, Ramachandran G. Polarization of line 
radiation in the presence of external electric quadrupole and 
uniform 
magnetic fields: II. Arbitrary orientation of magnetic field. JQSRT
2005;90:343-366.
\bibitem{rose} Rose ME. Elementary Theory of Angular Momentum. New 
York:
John Wiley, 1957.
\bibitem{hanle} Hanle W. \"Uber magnetische Beeinflussung der 
Polarisation der Resonanz fluoreszenz. Z Phys 1924;30:93-105.
\bibitem{BJ} Bransden BH, Joachain CJ. Physics of Atoms and 
Molecules. 
London: Longman, 1993.
\bibitem{shore} Shore BW, Menzel DH. Principles of Atomic Spectra. 
New 
York: John Wiley \& Sons, 1968.
\bibitem{muha1} Muha GM. Exact solution of the NQR $I= 1$ eigenvalue
problem for an arbitrary asymmetry parameter and Zeeman field 
strength
and orientation. J. Chem. Phys. 1980;73:4139-40. 
\bibitem{muha2}Muha GM. Exact solution of the eigenvalue problem for 
a 
spin-3/2 system in the presence of a magnetic field. J. Magnetic 
Resonance
1983;53:85-102.  
\bibitem{jauch} Jauch JM, Rohrlich F. The Theory of Photons and 
Electrons.
New York: Addison-Wesley, 1955.
\bibitem{hamilton} Hamilton J. The Theory of Elementary Particles. 
Oxford, 1959.
\bibitem{eugen} Eugen Merzbacher. Quantum Mechanics. USA, John Wiley 
$\&$ 
Sons Inc., 1970. 
\bibitem{weisskopf1} Weisskopf V, Wigner E. Berechnung der 
nat\"urlichen
Linienbreite auf Grund der Diracschen Lichttheorie. Z Phys 
1930;63:54-73.
\bibitem{weisskopf2} Weisskopf V, Wigner E. \"Uber die nat\"urliche 
Linienbreite in der Strahlung des harmonischen Oszillators. Z Phys 
1930;65:18-29.
Wiley
\bibitem{Mc} McMaster WH. Matrix representation of polarization. 
Rev. Mod. Phys.
1961; 33:8-28.
\bibitem{lanandlan}
Landi Degl'Innocenti M, Landi Degl'Innocenti E. An analytical 
expression 
for the Hanle-effect scattering phase matrix. Astron Astrophys 
1988;192:374-78.
\bibitem{obridko} Obridko VN. Scattering matrix for radiation in a 
magnetic 
field. Soviet Physics-Astronomy 1965;9:77-81.
\bibitem{ooth}
Yee Yee Oo. Studies in Astrophysical Line Formation Theory. PhD 
Thesis 2004. 
\end{thebibliography}
\end{document}